\documentclass[a4paper, 10pt,journal,twoside]{IEEEtran}
\usepackage[utf8]{inputenc}
\usepackage{eurosym}
\DeclareUnicodeCharacter{20AC}{\euro{}}
\usepackage{graphicx}
\usepackage{amsmath}
\usepackage{amssymb}
\usepackage{multirow}
\usepackage{algorithm}
\usepackage[noend]{algpseudocode}
\usepackage{makecell}
\usepackage[caption=false]{subfig}
\usepackage{stfloats}
\usepackage{float}
\usepackage{cite}
\usepackage{mathtools}

\usepackage{fancyhdr}

\usepackage{xr-hyper}
\usepackage{hyperref}

\usepackage[final]{changes}
\DeclarePairedDelimiter\floor{\lfloor}{\rfloor}

\makeatletter
\newcommand\floatc@myruled[2]{{\@fs@cfont #1} #2\par}
\newcommand\fs@myruled{\def\@fs@cfont{\bfseries}\let\@fs@capt\floatc@myruled
\def\@fs@pre{\hrule height.8pt depth0pt \kern2pt}%
\def\@fs@post{\kern2pt\vspace{-10pt}\hrule\relax}%
\def\@fs@mid{\kern2pt\hrule\kern2pt}%
\let\@fs@iftopcapt\iftrue}
\floatstyle{myruled}
\newfloat{Problem}{thbp}{prob}

\newcommand*{\addFileDependency}[1]{% argument=file name and extension
  \typeout{(#1)}
  \@addtofilelist{#1}
  \IfFileExists{#1}{}{\typeout{No file #1.}}
}
\makeatother

\newcommand*{\myexternaldocument}[1]{%
    \externaldocument{#1}%
    \addFileDependency{#1.tex}%
    \addFileDependency{#1.aux}%
}

\myexternaldocument{supplemental}

\begin{document}

\fbox{\begin{minipage}[b][1cm][c]{18cm}
\footnotesize This article has been accepted for publication in the Transactions on Intelligent Transportation Systems, IEEE. This is the author's version which has not been fully edited and content may change prior to final publication. Citation information: \url{https://dx.doi.org/10.1016/TITS.2022.316498}
\end{minipage}}

\author{Riccardo Iacobucci, Raffaele Bruno, Chiara Boldrini\thanks{The authors are with IIT-CNR, Via G. Moruzzi 1 -- 56124, Pisa, Italy,  e-mail: \emph{firstname.lastname}@iit.cnr.it.}
\vspace{-0.6cm}
}

% \markboth{IEEE Transactions on Intelligent Transportation Systems, ~Vol.~X, No.~Y, July~2017}%
% {Boldrini \MakeLowercase{\textit{et al.}}: Design and Performance Analysis of Fleet Relocation Strategies for Car Sharing Systems using Stackable Electric Vehicles}

\title{A Multi-stage Optimisation Approach to Design Relocation Strategies in One-way Car-sharing Systems with Stackable Cars}
% scalable modular approach for fleet relocation in shared mobility systems
\maketitle

\begin{abstract}
    One of the main operational challenges faced by the operators of one-way car-sharing systems is to ensure vehicle availability across the regions of the service areas with uneven patterns of rental requests. Fleet balancing strategies are required to maximise the demand served while minimising the relocation costs. However, the design of optimal relocation policies is a complex problem, and global optimisation solutions are often limited to very small network sizes for computational reasons. In this work, we propose a multi-stage decision support system for vehicle relocation that decomposes the general relocation problem into three independent decision stages to allow scalable solutions. Furthermore, we adopt a rolling horizon control strategy to cope with demand uncertainty. Our approach is highly modular and flexible, and we leverage it to design user-based, operator-based and robotic relocation schemes. Besides, we formulate the relocation problem considering both conventional cars and a new class of compact stackable vehicles that can be driven in a road train. We compare the proposed relocation schemes with two recognised benchmarks using a large data set of taxi trips in New York. Our results show that our approach is scalable and outperforms the benchmark schemes in terms of quality of service, vehicle utilisation and relocation efficiency. Furthermore, we find that stackable vehicles can achieve a relocation performance close to that of autonomous cars, even with a small workforce of relocators.
\end{abstract}


\section{Introduction}
\IEEEPARstart{M}{any} experts agree that we are at the dawn of a revolution in the automotive industry, which is driven by technological advances, digitalisation of mobility services, changes in people's mobility behaviours, as well as their perspective towards car ownership and the environmental impact of transportation systems~\cite{mckinsey_automotive_2016,bauer_cost_2018}. In particular, two global trends in the urban mobility landscape are particularly relevant to this study: $i)$ the increasing popularity and diffusion of shared mobility solutions, and $ii)$ the emergence of specialised vehicle concepts that attempt to reduce the \emph{road footprint} (namely, public space that is occupied by cars), providing more convenient personal urban mobility~\cite{mitchell_reinventing_2010}.

A wide range of different shared and on-demand mobility services, especially in dense urban environments, have emerged to enable users to gain short-term access to transport on an ``as-needed'' basis~\cite{shaheen_shared_2015}. One of the most prominent examples of these new on-demand mobility solutions is the \emph{one-way car-sharing} scheme. Members of such systems can rent a shared-used vehicle from a fleet operated by a private company or a public entity for one-way short trips, typically using a web-based or mobile app~\cite{schwieger2015global}. One-way car-sharing systems can be further categorised as station-based or free-floating schemes if the vehicle can be picked up and dropped off only from designated stations or at any location in the service area, respectively. 

A critical operational challenge for one-way car-sharing systems is how to ensure that there are sufficient available vehicles in each station (or across the regions of the service area) to satisfy the current and future rental requests. Indeed, it is well known that the distribution of a car-sharing fleet gets temporally and spatially imbalanced due to uneven demand patterns~\cite{barth_user-based_2004,boldrini_data_mining_2019}. The most straightforward approach to ensure the system balance is to over-dimension the fleet and station capacity to absorb demand fluctuations~\cite{george_fleet_2011}. A more effective approach for the car-sharing operator is to relocate empty vehicles where they are most needed, based on forecasted vehicle demand or short-term bookings. In particular, vehicle relocation can be performed: $i)$  by the users themselves, who are incentivised to carpool or to choose another trip destination; or $ii)$ by dedicated drivers, which is currently more common~\cite{illgen_review_relocation_2019}. There is a trade-off between the revenue loss due to lost rental requests and the costs associated with relocation operations, namely, fuel costs for the vehicle travelling empty, personnel cost for operator-based schemes~\cite{boyaci_integrated_2017}, or fare discounts for user-based relocation schemes~\cite{febbraro_user_based_2018}. 

There is abundant literature on the design of optimal relocation policies under different operational and business constraints (see Section~\ref{sec:relatedwork} for a detailed review). Typically, existing optimisation models are based on queuing theory~\cite{pavone_robotic_2012}, stochastic optimisation frameworks~\cite{nair_stochastic_2011}, mixed-integer linear programming (MILP)~\cite{boyaci_optimization_2015}, or model predictive control (MPC) tools~\cite{zhang_model_2016}. However, such models are computationally expensive and, for this reason, exact solutions are limited to small problem sizes. Moreover, most of the existing approaches develop the relocation model under the assumption that the time-ordered sequence of pickups and drop-offs in the car-sharing system (namely, the traffic demand) is known. However, deterministic approaches may be ill-suited to consider demand uncertainties that occur in real systems. On the other hand, stochastic optimisation problems based on probability distributions of demand are difficult to solve, and they require sophisticated heuristic algorithms, such as in~\cite{nair_stochastic_2011}.  

To cope with the intractability of finding an optimal global solution to the relocation problem for reasonable problem sizes and over a large time horizon, in this study we propose a novel \emph{modular} and \emph{multi-stage} decision-making tool for fleet rebalancing in one-way car-sharing systems. We show how to split the general relocation problem into three independent decision stages that can be solved sequentially. We will show that this approach scales well to very large instances of the relocation problem. The first decision stage focuses on the assessment of fleet imbalance; the second one on the selection of redistribution flows and routes between stations with over-accumulated vehicles to stations that experience vehicle shortages; and the third one on the scheduling of relocation tasks. Our optimisation framework relies only on short-term predictions of localised vehicle surpluses and deficiencies (also known as inventory imbalance). This approach mitigates the impact of demand uncertainties and dynamic traffic conditions on operational efficiency. Furthermore, it adopts a \emph{rolling} horizon control strategy to adjust relocation decisions to the evolving system state. 

\begin{figure}[t]
\begin{center}
\centering
\includegraphics[scale=0.2]{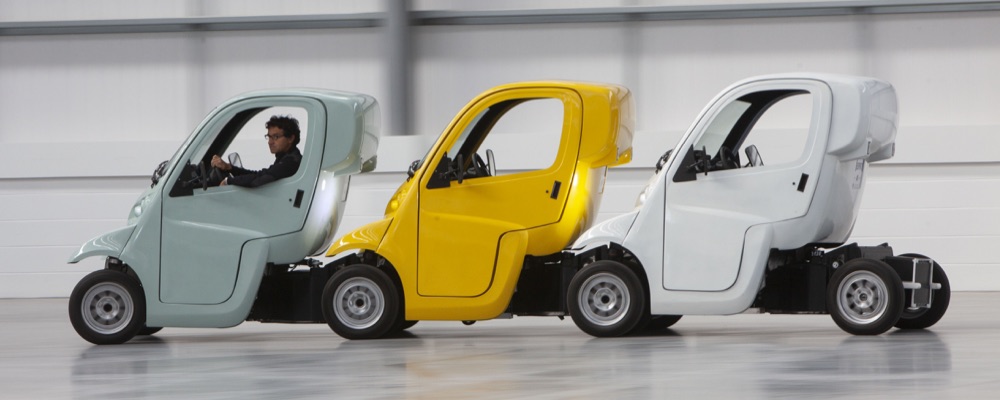}
\caption{The ESPRIT train of vehicles. }
\label{fig:esprit}
\end{center}\vspace{-15pt}
\end{figure}

The flexibility and generality of the proposed optimisation framework are demonstrated in two ways. First, we leverage our modular design to specify operator-based, user-based and robotic relocation policies that maximise the satisfied demand while minimising the time a vehicle spends in relocation operations. Second, we formulate our optimisation models not only for conventional cars, but also for a class of \emph{stackable electric cars} that can be coupled together when parked (to save space), and driven together as a ``road train''. The design of compact electric cars that can be folded and stacked in line, which are intended to be used for short-distance urban trips in car-sharing systems, is not a novel concept, being the MIT CityCar\footnote{\url{https://en.wikipedia.org/wiki/CityCar}.} and the Hiriko car\footnote{\url{https://en.wikipedia.org/wiki/Hiriko}.} the most famous examples. More recently, this concept has been further developed and also extended to include coupling capabilities that allow the creation of a train of vehicles than can be driven (and even recharged) together. An example of such concept is the ESPRIT car (illustrated in Fig.~\ref{fig:esprit}), which has been prototyped in 2018 by a European consortium\footnote{\url{http://www.esprit-transport-system.eu/}.}. One important feature of the ESPRIT prototype is semi-autonomous towing capabilities, whereby each vehicle, when being parked at a station, is stacked in an automated way to the vehicles already parked. Preliminary studies have shown that ESPRIT cars could be particularly useful for car-sharing systems, as they would enable more efficient redistribution of the fleet (one driver can relocate two or more vehicles~\cite{boldrini_relocation_2017}), as well as more efficient utilisation of the charging infrastructure (one charging station could serve multiple vehicles simultaneously~\cite{biondi_power_sharing_2016}). There are similarities between the relocation problem for stackable cars and the relocation problem in bike-sharing systems, which typically use trucks to relocate simultaneously a large amount of bicycles~\cite{raviv_optimal_2013,bulhoes_static_2018}. However, stackable cars offer more flexibility as they can be relocated without using a dedicated fleet of service vehicles. 
%It is interesting to point out that 
An operator-based relocation strategy using dedicated towing vehicles was also experimented in the past in the car-sharing system deployed at UCR, called Intellishare~\cite{barth_intellishare_2003}. However, manual towing to relocate vehicles is very inefficient and slow. To the best of our knowledge, this paper is the first to formalise the relocation problem in a car-sharing system by considering stackable cars that have \emph{semi-autonomous towing capabilities}, and to apply this concept to both operator-based and user-based policies.

 To assess the efficiency and capabilities of the proposed framework, we tested different relocation techniques, using taxi trip data from New York City in 2018. The data set contains pickup and drop-off times/locations of more than 200.000 trips per day, offering a large-scale instance of the relocation problem. We also compare our relocation schemes with two state-of-the-art solutions, one designed for station-based one-way car-sharing systems~\cite{boyaci_optimization_2015}, and one designed for free-floating bike-sharing systems~\cite{caggiani_modeling_2018}. Trade-offs between the level of service offered, fleet size, relocation efficiency, model complexity are discussed. Our results show that our approach ensures lower computation times than the considered benchmark schemes, as well as improved quality of service, vehicle utilisation and relocation efficiency. Furthermore, we demonstrate that a small team of drivers relocating stackable cars is sufficient to approach the efficiency of a shared-used fleet of self-driving vehicles.

The remainder of this paper is organised as follows. Section~\ref{sec:relatedwork} reviews related work and further elaborates on the novelty of the proposed approach. Section~\ref{sec:decision} introduces the system model and the proposed methodology. Section~\ref{sec:decision} presents the relocation models. Section~\ref{sec:evaluation} compares the proposed relocation techniques with two benchmark schemes, while Section~\ref{sec:conclusions} discusses the research conclusions and provides recommendations for future research.

%
%---------------------------------------------------
\section{Related Work\label{sec:relatedwork}}
\noindent
The fleet balancing problem in one-way car-sharing systems and, more in general, in shared mobility systems, has been extensively studied in recent years. The reader is referred to~\cite{boyaci_optimization_2015,illgen_review_relocation_2019,ho_DRP_survey_2018} for a comprehensive review. Given the focus of our study, this overview concentrates on the \emph{proactive} relocation problem, where vehicles are dispatched to certain stations or service zones to satisfy future predicted demand. 
 
% QUEUES THEORY 
 
Two main control approaches are adopted for proactive relocation. A first approach leverages a stochastic representation of traffic demands. One class of solutions leverages stochastic fluid models, which describe the car-sharing system as a queuing network~\cite{pavone_robotic_2012,meyer_toward_2014,zhang_control_2016}. These models assume that the demand pattern can be simply represented as a set of aggregated flows of vehicles between network nodes, and modelled as a Poisson arrival process. Moreover, queuing theory is used to model available vehicles and waiting customers in the system. Then, average relocation rates are determined by finding the optimal steady-state solution of the system dynamics. However, under time-varying demand conditions, a steady state might not exist. Moreover, this modelling approach is not readily adaptable to varying levels of demand over time. A second class of solutions developed stochastic programming models for fleet rebalancing with uncertain demands that are described through probability functions~\cite{nair_stochastic_2011,2019_arxiv_stochastic}. However, these models are typically solved using heuristic approaches or Monte Carlo methods. Dynamic operator-based relocation strategies are presented in~\cite{repoux_dynamic_2019} for a station-based car-sharing system with short-term reservations and limited station size. The first strategy is based on the requirement to have at least one vehicle at each station at any time, while a second strategy is based on a Markov model to estimate future shortage of vehicles or parking spots at stations.

% SPACE TIME NETS

A second approach relies on the construction of space-time networks to describe system dynamics with individual rental requests. Then, vehicle relocation is typically formulated as a mixed-integer linear programming (MILP) problem \cite{kek_relocation_2009,weikl_relocation_2013,boyaci_optimization_2015,zhang_model_2016,boyaci_integrated_2017,wang_relocation_2019}. Specifically, the authors in~\cite{kek_relocation_2009} formulate a MILP problem to minimise the total generalised cost of relocation, taking into consideration movement and relocation costs, staff cost and penalty costs of rejected rental requests. The branch-and-bound technique is used to find exact solutions to the problem for a small dataset containing less than 2000 trips. Relocation strategies for free-floating systems are proposed in~\cite{weikl_relocation_2013} by combining a macroscopic relocation optimisation policy of moving vehicles between zones, with a rule-based heuristic for the relocation of individual vehicles. In~\cite{repoux_dynamic_2015} a rolling horizon optimisation approach is proposed to maximise the number of relocations performed over a short period of time ahead considering both in advance reservations and last-minute trip requests. In~\cite{boyaci_optimization_2015} an integrated optimisation framework is developed to model operator-based relocation, station planning and fleet sizing jointly to maximise the net revenue of the car-sharing operator. The work does not take into account operator rebalancing, but it only guarantees that the total time spent to relocate vehicles does not exceed the total available working hours for a shift of the relocation personnel. Vehicle and personnel relocation in a car-sharing system with reservations is modelled in~\cite{boyaci_integrated_2017}. A model predictive control approach with dynamic continuous relocation is presented in \cite{zhang_model_2016}. The main limitation of these works is that the MILP formulations are computationally very demanding as they explicitly model the routing of vehicles in the system, i.e. which route a vehicle should take when moved by a relocator or rented by customers. Thus, either very small systems are tested, or approximate algorithms (e.g., aggregated models with relaxed constraints~\cite{boyaci_optimization_2015}, station clustering~\cite{weikl_relocation_2013,boyaci_integrated_2017}, or zoning schemes~\cite{wang_relocation_2019}) are proposed to determine sub-optimal solutions in large-scale scenarios. Furthermore, the creation of a time-space model of the state of the car-sharing system requires a fixed and \emph{deterministic} demand pattern (e.g. the time-ordered  sequence  of  pickups  and  drop-offs  in  the  car-sharing  system). This approach is ill-suited to consider demand uncertainties that occur in real systems.

% USER BASED RELOC

A few studies deal with user-based relocation, where users are encouraged and compensated for changing their behaviour to relocate vehicles \cite{pfrommer_dynamic_bike_2014,febbraro_user_based_2018}. \added{While these works show that user-based relocation can increase the system profitability and the served demand, the system performance significantly depends on the level of spatial and temporal flexibility of the demand \cite{2019_boyaci_flexibility}}. A \added{potential} advantage of stackable vehicles is that user-based relocation may be possible without changing trip patterns just by appending extra vehicles to specific passenger trips \cite{boldrini_stackable_2017,boldrini_relocation_2017}. 

% RELOCATOR'S RELOCATION

%It is also important to point out that 
Operator-based relocation schemes suffer from the intrinsic complexity of optimising the routing and rebalancing of operators, as these too become unbalanced~\cite{smith_rebalancing_2013}. In practice, the relocation of operators between tasks is generally accomplished through a second operator in a service car, or by having operators move among regions by other means, such as public transport or by a folding bike which can be stowed in the relocating vehicle \cite{illgen_review_relocation_2019}. Both approaches have limitations: the first one forces to double the number of relocators, thus increasing relocation costs. The second approach requires longer times for the relocation of operators between tasks. A hybrid approach is to have relocators drive vehicles with passengers when relocating themselves, akin to a mixed car-sharing/taxi service \cite{smith_rebalancing_2013}. 

% BIKE SHARING SYSTEMS

The car-sharing relocation problem is also related to relocation in bike-sharing systems, where a dedicated fleet of trucks is used to relocate multiple bikes at once. Most of the literature on bike-sharing relocation has addressed the rebalancing problem as a variant of the one-commodity pickup-and-delivery capacitated vehicle routing problem. Most studies have considered a static problem in which the changes in the bike usage rate are negligible during the repositioning period (i.e., night relocation)~\cite{raviv_bike_2013,li_static_bile_2016}. Other works consider the dynamic problem in which the usage rate varies over time (e.g. ~\cite{caggiani_dynamic_bike_12,pfrommer_dynamic_bike_2014}) or they allow the repositioning trucks to visit a station multiple times (e.g.~\cite{salazar-gonzalez_bike_2015}). Models for the relocation problem with multiple trucks and multiple visits to stations have also been developed~\cite{bulhoes_static_2018}, but they are computationally costly. A simplified and tractable method assumes single visits to stations and divides the problem into two separate vehicle routing problems: one for bike pick-up, and the following one for dropping off bikes and returning to the depot \cite{caggiani_modeling_2018}. To conclude, it is important to point out that bike-sharing systems generally assume few relocation operations during the day, since the cost of the operator/truck is much higher than the bikes themselves. Conversely, in a car-sharing system, the cost of the vehicles is one of the highest costs for the system, so a strategy more reliant on frequent relocations and a smaller fleet is favourable.

% our contributions

The optimisation model proposed in this paper differs in many respects from the papers mentioned above. In particular, we deal with the computational complexity of the problem by using a multi-stage modelling approach, rather than a unified global optimisation. Furthermore, our solution is also robust to demand uncertainties as it leverages a rolling horizon control approach. Finally, the developed relocation models are not restricted to conventional cars but are generalised to be also used with stackable cars. While bike sharing and car sharing with stackable cars have many similarities, especially at the system planning level, stackable cars offer additional features (e.g., vehicle coupling) and increased flexibility, which demand new model formulations.

%
%---------------------------------------------------
\section{A Multi-stage Decision Support System for Vehicle Relocation\label{sec:decision}}
\noindent
Before introducing the model formulation, we present the design principles and the modular architecture of our decision support system for fleet rebalancing in one-way car-sharing systems. For the sake of simplicity, let us assume that time is discretised into time slots. A vehicle relocation strategy is a function that takes as input the car-sharing system status in a given time slot and outputs the set of relocation actions for a number of future time slots during which the relocation process is to be optimised. A relocation action consists in the transfer of one or more vehicles from a zone/station where there is an accumulation of vehicles (called \emph{feeder}) to a zone/station where there is a shortage of vehicles (called \emph{receiver})\footnote{Note that our model formulation can be applied to both station-based and free-floating systems.}. Depending on the relocation strategy, vehicle relocation can be carried out by a professional driver, a user, or even autonomously if self-driving cars are available. 

\begin{figure}[ht]
\center
\includegraphics[trim={1.5cm 1cm 1.5cm 10cm},clip,angle=0,  width=\columnwidth]{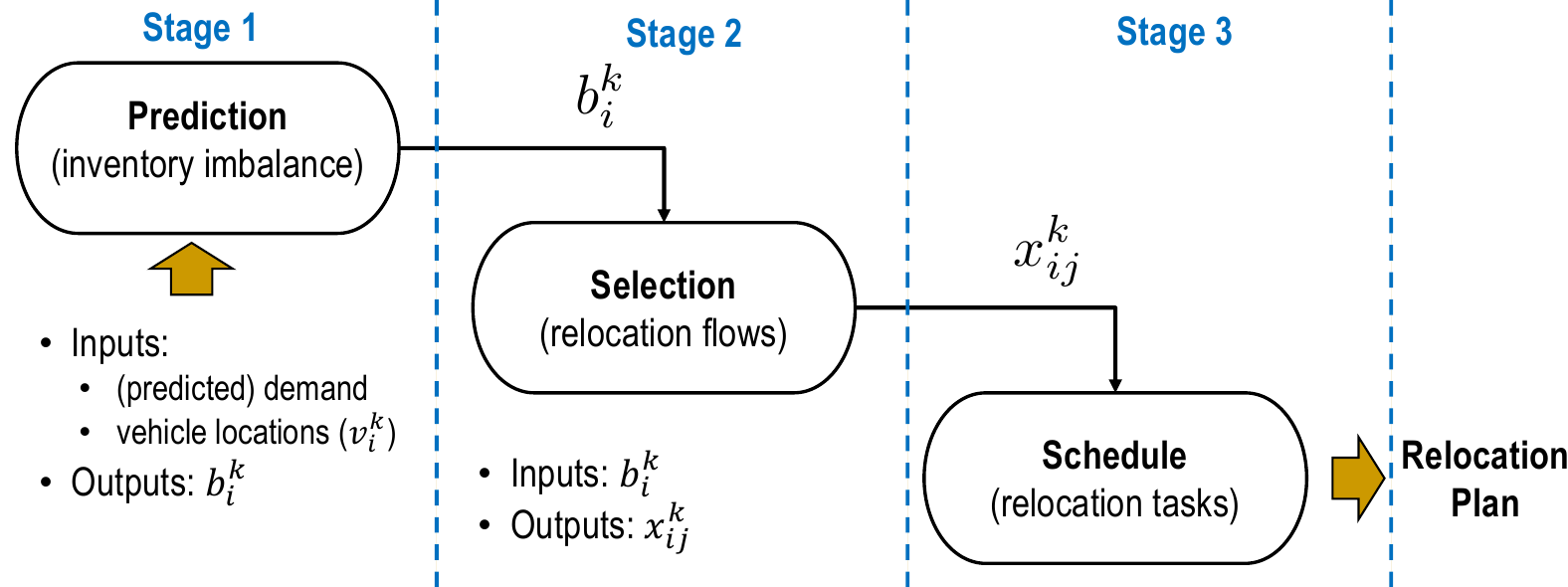} 
\caption{Functional architecture of the multi-stage decision support system for vehicle relocation we propose in this study. The blocks in the diagram represent the main components of the systems with the input and output variables. Variable $v^k_i$ and $b_{i}^{k}$ denotes the available vehicles and the vehicle inventory imbalance, respectively, of station $i$ at the decision point $k$, while $x_{ij}^{k}$ is the relocation flow from station $i$ to station $j$ during the $k$-th decision interval. The formal definition of these variables is provided in Section~\ref{sec:relocation}.}
\label{fig:decision_chain}
\vspace{-0.2cm}
\end{figure}

As discussed in Section~\ref{sec:relatedwork}, for practical size problems solving an optimisation model for the selection of individual relocation actions over an infinite control horizon may be a computationally intractable task. To address this issue, in this study we propose a \emph{problem decomposition} approach that splits the original relocation problem into three simpler sub-problems that can be solved sequentially to determine a near-optimal (or at least well-performing) relocation policy. The structure of this approach is presented in Fig.~\ref{fig:decision_chain}. 

\textbf{Prediction module}: The first stage of the decision chain consists in predicting the maximum inventory imbalance (i.e. surplus or lack of vehicles) of each zone in the car-sharing operational area during a future time interval. Then, a feeder is any region (or station) with a positive imbalance, i.e., a foreseen surplus of vehicles. At the same time, a receiver is any region (or station) with a negative imbalance, i.e., a foreseen deficiency of vehicles. This component takes as \emph{input} the current inventory levels of the zones (i.e., how many vehicles are currently available) and the future demand patterns, and it provides as \emph{output} the estimate of the vehicle inventory imbalance of each station. A precise characterisation of the imbalance level is difficult due to the uncertainty of future demands~\cite{miao_19_demand_uncertainty,boldrini_data_mining_2019}, and the complex interactions between the demand processes of different zones ~\cite{raviv_optimal_2013}. The design of our predictors is detailed in Section~\ref{sec:prediction}.

\textbf{Selection module}: This component of our decision-support system takes as \emph{input} the inventory imbalance of each zone and provides as \emph{output} the total number of vehicles to move between each feeder-receiver pair. The amount of time the car-sharing operator looks in the future to determine the relocation flows defines the planning horizon of the redistribution process. As discussed in Section~\ref{sec:relatedwork}, long-term planning horizons (e.g., full operating days) typically assume known and static demand (e.g., a reservation-based system, where all the customer requests must be performed in advance), or a nearly idle system (e.g. when relocation is carried out during the night) ~\cite{gambella_optimal_2018}. On the contrary, short-term planning horizons are more suitable to cope with uncertain demands and to continuously exploit feedback about redistribution efficiency. Finally, different objective functions can be defined for the relocation process. There is a cost to move a vehicle between zones, and this cost typically depends on the distance between zone centroids. The model formulation of the second decision stage is detailed in Section~\ref{sec:relocation_flows}. 

\textbf{Schedule module}: The third and final stage of the decision chain takes as \emph{input} the set of relocation flows and schedules the sequence of relocation tasks that implement those relocation flows. A relocation task defines the relocation operations for individual (or group of) vehicles, namely how to split the aggregated relocation flow into a sequence of relocations of individual vehicles or trains of vehicles. Intuitively, feasible relocation tasks depend on the specific relocation technique. For instance, in case of an operator-based scheme, if the driver is not already available at the selected feeder, the car-sharing operator has to send one from another zone, incurring additional costs and delays. On the other hand, in a user-based scheme, an incentive (monetary or otherwise) should be offered to customers to contribute to vehicle rebalancing. Moreover, the time before a customer willing to relocate a vehicle arrives at a given feeder zone is uncertain. 

A key advantage of the multi-stage decision process in Fig.~\ref{fig:decision_chain} is that different relocation techniques can be easily plugged into the system by only adapting the last stage of the decision chain. Least-cost scheduling algorithms for different relocation techniques, both for conventional, stackable and robotic cars are developed in Section~\ref{sec:relocation_task}. A pseudo-code of the implementation of the full relocation algorithm that is executed by the proposed multi-stage decision support system is also provided in Appendix~A of the supplemental material.
%

%
%---------------------------------------------------
\section{Relocation Models\label{sec:relocation}}
\noindent
In this section, we present the mathematical formalisation of the proposed relocation models. In Section~\ref{sec:system}, we first define the sets and indices used to describe the model, as well as the functions, variables and parameters. Then, we describe in details the models for the three optimisation stages of the relocation policy presented in Section~\ref{sec:decision}.   
%
%
%---------------------------------------------------
\subsection{Problem definition\label{sec:system}}
\noindent
Let us assume that the operational area of the car-sharing system is partitioned into $N$ zones. Let $\mathcal{S}=\{s_1,s_2,\ldots,s_N\}$ be the set of zones. We also assume that there is an unlimited number of parking stalls reserved for shared vehicles within each service zone. Then, a demand pattern is associated with a zone, which represents an aggregation of pickup and drop-off events of vehicles within that zone. For the sake of simplicity, and without loss of generality, we cluster together the origin and destination locations of a car-sharing trip into the centroid of the related zone\footnote{The same formalism can be readily applied to a station-based car-sharing system by substituting the zone centroids with the station locations.}. Then, the travel time $T(i,j)$ between centroids of zones $s_i$ and $s_j$ is constant and derived from historical traffic data. Without ambiguity, in the following, we use the index $i$ to refer to zone $s_i$. Finally, in our model, we assume that customers do not wait for a vehicle, nor they change departure or arrival zones. In other words, a request for a car-sharing trip departing from a zone $i$ with destination $j$ (with the possibility that $j \!=\! i$) is admitted only if an empty vehicle is available in zone $i$. Otherwise, the user leaves the system.  

\begin{figure}[tbhp]
    \centering
    \includegraphics[trim={1cm 1.2cm 8cm 12.5cm},clip,angle=0,width=1.0\columnwidth]{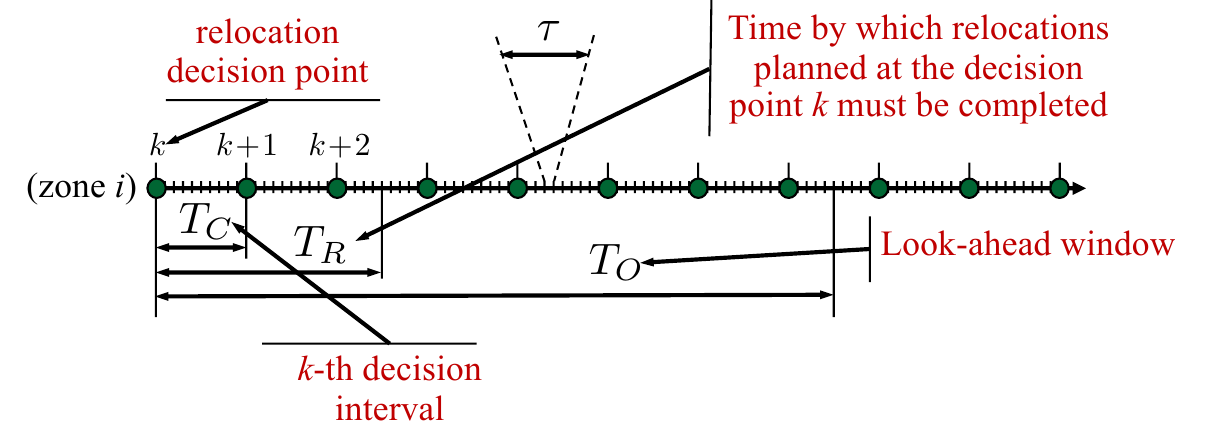}
    \caption{Illustrative timeline of the relocation decision process with $n_C=10$, $n_R=25$ and $n_O=75$ time slots. Circles represent the time points when relocation decisions are taken}.
    \label{fig:timeline}
    \vspace{-0.2cm}
\end{figure}

For computational efficiency, time is discretised into times slots of duration $\tau$. Then, all the time variables of the model are expressed as multiples of this time unit. Similarly to~\cite{gambella_optimal_2018}, we adopt a \emph{rolling horizon} approach to decide the relocation plan. Specifically, each operation day is split into time periods of duration equal to $n_C$ time slots (i.e. $T_C = n_C \tau$), called \emph{planning periods}. Then, the relocation plans are computed at the beginning of each planning period using information about vehicle locations, rental requests, and surpluses/deficiencies of vehicles at each zone. Our relocation model assesses the effect of relocation decisions on the imbalance of vehicle supply considering a look-ahead time window of $n_O$ time slots. In other words, $T_O=n_O \tau$ is the model \emph{prediction time horizon}, with $n_O > n_C$ (see Fig.~\ref{fig:timeline}). Following the rolling horizon approach, the relocation decision process is iterated over the subsequent planning periods. 
When the time horizon rolls from the decision point k to the decision point $(k \!+\!1)$, the predicted state of the car-sharing system for the next $n_O$ time slots is updated, (see Section~\ref{sec:prediction} for the details about the prediction methods). Note that when the system state is updated at time $(k \!+\!1)$, the decision taken at the decision point $k$ and not yet completed may not be optimal anymore, \added{according to the state information and updated trip information that is available at decision point $(k \!+\!1)$. Typically, in a rolling-horizon control approach, control decisions taken in a previous stage of the decision process, and not yet completed when the following decision stage starts, should not be modified \cite{scattolini_architectures_2009}. Thus, any pending (i.e., not yet finished or started) relocation task from decision period $k$ is not modified when a new relocation plan is computed at time $(k \!+\!1)$ (see Fig. \ref{fig:scheme} for an illustration of a pending relocation task). It is important to point out that} the use of short-term demand predictions and \added{ decision periods shorter than the typical duration of a relocation trip,} help to cope with demand variability.

Without loss of generality, we assume that a relocation task starts at the beginning of a time slot. However, relocation operations (i.e., the physical redistribution of vehicles) can generally last more than one time slot. Furthermore, a relocation decision can entail multiple relocation operations from the same pair of feeders and receivers. However, we require that the sequence of consecutive relocation operations between a feeder-receiver pair that follow the decisions calculated at time $k T_C$ are completed within a time $T_R = n_R \tau$, with $n_C \le n_R \le n_O$. A bound on the maximum time to complete a relocation task is beneficial to limit the relocation scope and to avoid relocating vehicles between very distant stations. Furthermore, relocation operations must finish within the model prediction time horizon to ensure that only the latest predicted state is used when calculating the relocation decision. Finally, setting up $T_C \le T_R$ reduces the probability that the system is idle because all relocation tasks are completed before a new decision interval starts. For the sake of clarity, Fig.~\ref{fig:timeline} illustrates an example of the relation between $\tau$, $T_C$, $T_O$ and $T_R$.   
%
%
%---------------------------------------------------
\subsection{Prediction of inventory imbalance\label{sec:prediction}}
\noindent
As defined in Section~\ref{sec:system}, at the $k$-th decision interval, the relocation policy needs to update the  information about the inventory imbalance of each zone in the future $n_O$ time slots to decide, first, which zones can be feeders and receivers and, second, how many vehicles should be relocated. Let $b^k_i$ denote the expected inventory imbalance for zone $i$ during interval $[k T_C,k T_C + T_O]$ (see Fig.~\ref{fig:timeline}) in case no new relocations are scheduled at time $k T_C$. Ideally, to exactly compute the $b^k_i$ value we would need the in-advance knowledge of the functions $A^k_i(t)$ and $C^k_i(t)$, with $t =1,\ldots,n_O$, to model the vehicle arrival and departure processes of zone $i$ in a deterministic manner\footnote{Clearly, in a real-world system $A^k_i(t)$ and $C^k_i(t)$ can only be predicted.}. More formally, let us assume that $A^k_i(t)$ provides the exact number of vehicles (either driven by a relocator or a customer) that arrive in zone $i$ during time slot $t$ in $[k T_C,k T_C + T_O]$. Similarly, let us assume that $C^k_i(t)$ provides the exact number of customers requesting a shared vehicle from zone $i$ during time slot $t$ in $[k T_C,k T_C + T_O]$. Let $v_i^k$ be the number of already available vehicles within zone $i$ at the beginning of the the $k$-th decision interval (a.k.a. the initial inventory level of zone $i)$. Now, let us introduce the variable $I^k_i(t)$, which measure the \emph{virtual} inventory level of zone $i$ at time slot $t$, namely the number of vehicles in zone $i$ at time slot $t$ if all customer trip requests would be accepted. It holds that $I^k_i(t)$ is given by:
\begin{equation}
I^k_i (t) = v_i^k + \sum_{n=1}^{t}  \left [A^k_i(n)-C^k_i(n) \right ]\; . \label{eq:I_i}
\end{equation}

\begin{figure}[ht]
    \centering
    \subfloat[][]{ % [trim=left bottom right top, clip]
    \includegraphics[trim={2cm 1cm 15cm 12cm},clip,angle=0,width=0.45\columnwidth]{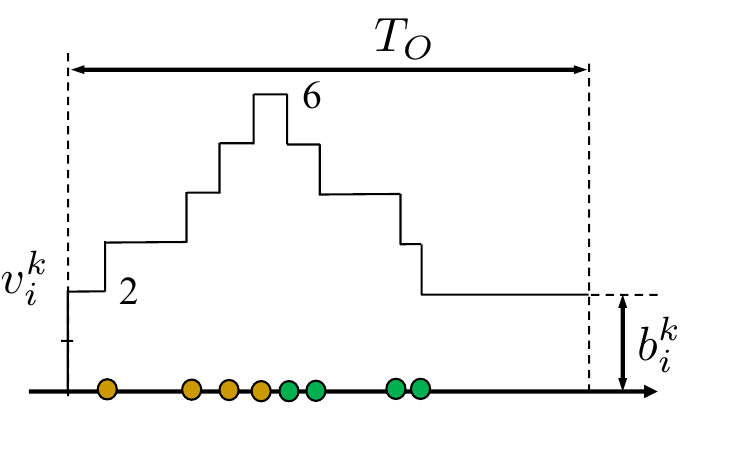}
    \label{fig:inventory_feeder}
    }
    \subfloat[][]{
    \includegraphics[trim={2cm 1cm 15cm 13cm},clip,angle=0,width=0.45\columnwidth]{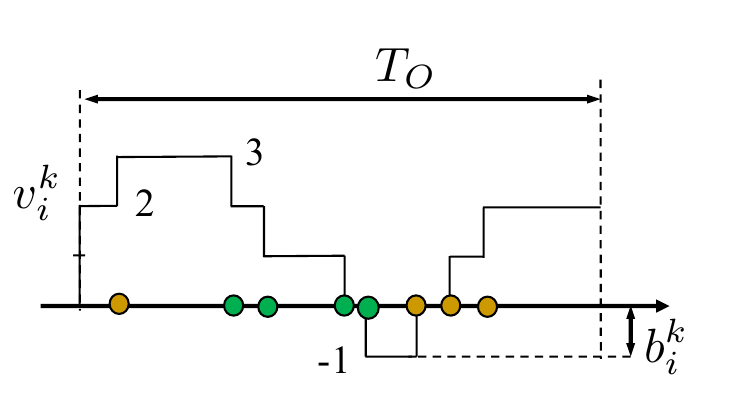}
    \label{fig:inventory_receiver}
    }
    \caption{Examples of the virtual inventory level of a (a) feeder station and a (b) receiver station. Orange circles denote vehicle arrivals, while green circles denote customers' trip requests. Initial inventory level $v^k_i=2$. Both cases include four vehicle arrivals and four customers' trip requests in total, but different time orders.
    }
    \label{fig:inventory}
\vspace{-0.3cm}
\end{figure}

For the sake of clarity, Fig.~\ref{fig:inventory} shows two illustrative examples of the temporal evolution of the $I^k_i (t)$ function for a zone $i$ with an initial inventory level equal to two vehicles.
Owing to the definition of $I^k_i (t)$, the inventory imbalance of zone $i$ during interval $[k T_C,k T_C + T_O]$ can be computed as:
\begin{equation}
b^k_i = \min\limits_{t=1,\ldots,n_O} I^k_i (t) \label{eq:bi_exact}
\end{equation}
A positive inventory imbalance quantifies the maximum number of vehicles that can be removed from zone $i$ ensuring that $I^k_i (n) \ge 0$ throughout time interval $[k T_C,k T_C + T_O]$ (see Fig.~\ref{fig:inventory_feeder}). Similarly, a negative inventory imbalance quantifies the minimum number of vehicles that should be moved to zone $i$ to ensure that $I^k_i (n) \ge 0$ throughout time interval $[k T_C,k T_C + T_O]$ (see Fig.~\ref{fig:inventory_receiver}). Then, the set $\mathcal{F}^k$ of feeder zones at the $k$-th decision interval is simply given by $\mathcal{F}^k=\{s_i \in \mathcal{S} | b^k_i > 0\}$. Similarly, the set $\mathcal{R}^k$ of receiver zones at the $k$-th decision interval is given by $\mathcal{R}^k=\{s_i \in \mathcal{S} | b^k_i < 0\}$.

As shown in Equation~\ref{eq:bi_exact}, a precise estimate of the $b^k_i$ variable would require to know the \emph{time-ordered} sequence of vehicles and customers arrivals at each zone. Unless a strict reservation-based system is employed, car-sharing operators typically have an uncertain knowledge of $A^k_i(t)$ and $C^k_i(t)$, mainly leveraging historical data for rental requests. Depending on the quality and granularity of available data, different approaches can be devised to estimate $b^k_i$, and two of them are discussed in the following.   
%
%
%---------------------------------------------------
\subsubsection{Worst-case estimate of \texorpdfstring{$b^k_i$}{b}\label{sec:worst-case_estimate}}
\noindent
To obtain a worst-case estimate of the $b^k_i$ parameter, the car-sharing operator can leverage \emph{average} estimates of the demand over the observation interval $T_O$. Specifically, let us assume that the car-sharing operator knows: $i)$ the total number $C^k_i$ of expected customers' requests for trips departing from zone $s_i$ in $[k T_C,k T_C + T_O]$; $ii)$ the average number $D^k_i$ of passenger trips that start after time $k T_C$ and terminate at zone $i$ in $[k T_C,k T_C + T_O]$; and $iii)$ the number $R^k_i$ of vehicles that are currently en route to zone $i$, including empty vehicles and vehicles with passengers, and that are expected to arrive before $kT_C + T_R$. Intuitively, a rough estimate of $b_i^k$ at each zone could be given by the net balance of vehicles arriving, vehicles leaving, and vehicles that are already parked there. Among these components, the estimate of $D^k_i$ is typically the most unreliable, as it depends on the complex interactions between the demand processes of different zones~\cite{raviv_optimal_2013}. Thus, a conservative estimate of $b_i^k$ could ignore the contribution of $D^k_i$. Note that this does not affect the quality of service perceived by the customers: if the $D^k_i$ vehicles do indeed arrive in the end, customers will experience a greater availability in zone $i$. 

Based on the above considerations, it is easy to observe that the worst-case estimate of $b^k_i$ is given by  
\begin{equation} % manca D, veicoli che arrivano tra [kTc, kTc+Tr]
b_i^k = v_i^k + R_i^k - C_i^k  \; . \label{eq:worst-case_imbalance} % D??
\end{equation}
Basically, Eq.~\eqref{eq:worst-case_imbalance} assumes that all user requests are generated before new vehicles are dropped off at a given zone. 
%!!!COMMENT!!!
%\textbf{NON E' DETTO PERCHE' IN FIGURA NON SAPPIAMO QUALI ARRIVANO PRIMA DI $kT_C + T_R$!!}
% Credo che il tuo commento derivi dal fatto che la definizione di A^k_i(t) non era chiara. Ho spiegato meglio che A^k_i(t) specifica i veicoli che arrivano durante un slot sia se sono portati dai clienti che dai relocator. In questo modo il commento non dovrebbe essere più valido
% bik in fig. 4 non è definibile in quanto non è definito R_ik che dipende da T_R
%
%
%---------------------------------------------------
\subsubsection{Probabilistic estimate of \texorpdfstring{$b^k_i$}{b}\label{sec:probabilistic_estimate}}
We can obtain a probabilistic estimate of the $b^k_i$ parameter if we assume that the car-sharing operator at least knows the \emph{demand probability distributions} of each zone. More formally, let $f^{k,t}_A(n;i)$ and $f^{k,t}_C(m;i)$ denote the probabilities that $n$ vehicles arrive at zone $i$ during time slot $t$ in $[k T_C,k T_C + T_O]$, and $m$ user requests for trips departing from zone $i$ arrive during time slot $t$ in $[k T_C,k T_C + T_O]$, respectively. Moreover, we assume that $f^{k,t}_A(n;i)$ is defined over the finite set $[0,\beta_V]$, with $\beta_V \in \mathbb{Z}_{> 0}$, while $f^{k,t}_C(n;i)$ is defined over the finite set $[0,\beta_C]$, with $\beta_C \in \mathbb{Z}_{> 0}$. In general, these demand probability distributions can be estimated by historical trip data using simple averaging (as explained in Section~\ref{sec:evaluation}), or more sophisticated methods, such as kernel density estimation (e.g. in~\cite{2019_arxiv_stochastic}) or ML techniques (e.g. in~\cite{boldrini_data_mining_2019}). Now, we can model the virtual inventory variable $I^k_i (t)$ as a time-varying Markov chain. To compute the transition probabilities we introduce the auxiliary random variable $\Delta^k_i(t) \in \mathbb{Z}$, denoting the increment of the virtual inventory level in time slot $t$, i.e., $\Delta^k_i(t) = [ A^k_i(t)-C^k_i(t) ] $. We remind that the inventory level is increased by one for each vehicle arrival and decreased by one for each rental request. For the sake of simplicity and computational efficiency, we require that users issuing a rental request during time slot $t$ can pick-up the vehicle only at the end of the time slot. In this case, it holds that:
\begin{align}
     \Pr\{\Delta^k_i(t) \!=\!  l \} & = \sum_{n=0}^{\beta_V}\sum_{m=0}^{\beta_C}\big [ f^{k,t}_A(n;i) - f^{k,t}_C(m;i)\big ] \delta_{(n-m),l} \\ \nonumber 
     & \qquad \textrm{ with  } l \in [-\beta_V,\beta_C]\; ,
     \label{eq:prob_imbalance}
\end{align}
where $\delta_{i,j}$ is the Kronecker delta function: 0 if $i \neq j$; 1 if $i=j$. Owing to the law of total probability, we can write:
\begin{equation}
    \Pr\{I^k_i (t+1) \!=\! z\}  \!=\!  \sum_{l=-\beta_V}^{\beta_C} \Pr\{I^k_i (t) \!=\! z-l\} \cdot \Pr\{\Delta^k_i(t) \!=\!  l \} \; ,
    \label{eq:pr_X}
\end{equation}
From a generic initial state $I^k_i (0)$, the inventory level evolves following paths that are constrained by the $\beta_V$ and $\beta_C$ parameters. More formally, let ${}^+z^{t}_i$ and ${}^-z^{t}_i$ be the maximum and minimum values such that  $\Pr\{I^k_i (t) = {}^+z^{t}_i\}$ and $\Pr\{I^k_i (t) = {}^-z^{t}_i\}$ are not null. It is straightforward to note that it holds:
\begin{align}
    {}^+z^{t}_i & =  I^k_i (0)+ \beta_V t  \label{eq:z+}\\
    {}^-z^{t}_i & =  I^k_i (0)-\beta_C t  \; . \label{eq:z-}
\end{align}
The rationale of Eqs.~\eqref{eq:z+} and~\eqref{eq:z-} is that after each time slot the maximum increment of the $I^k_i (t)$ random variable is $\beta_V$, while $\beta_C$ is the maximum decrement. 

Now we can define a methodology to provide a probabilistic estimate of the inventory imbalance during interval $[k T_C,k T_C + T_O]$ . Let $F^k_i(v^k_i)$ denote the total probability of observing a negative $I^k_i (t)$ value over the period $T_O$ if $I^k_i (0)= v^k_i$. It holds that:
\begin{align}
   F^k_i(v^k_i) & = \sum_{t=1}^{n_O} \sum_{z={}^-z^{t}_i}^{-1} \Pr\{I^k_i (t) = z | I^k_i (0)= v^k_i\}  \; .
\end{align}
Intuitively, $F^k_i(v^k_i)$ represents the probability that there is shortage of vehicle at zone $i$ during $[k T_C,k T_C + T_O]$, given a demand probability distribution. Thus, $F^k_i(v^k_i) >0 $ implies that zone $i$ is a receiver and the inventory imbalance $b^k_i$ is the smallest negative value that ensures that $F^k_i(v^k_i+b^k_i) \le \epsilon$, with $\epsilon$ small. On the contrary, $F^k_i(v^k_i) =0$ implies that zone $i$ is a feeder and the inventory imbalance $b^k_i$ is the smallest positive value that ensures that $F^k_i(v^k_i-b^k_i) \ge \epsilon$, with $\epsilon$ small.

In Appendix~B of the supplemental material we present a comparison of the two proposed methods and the evaluation of their impact on the relocation efficiency.
%
%
%
%---------------------------------------------------
\subsection{Selection of relocation flows \label{sec:relocation_flows}}
\noindent
In this section, we formulate an optimisation model to calculate the set of relocation flows between pairs of feeders and receivers that satisfy a two-fold objective: $1)$ to balance the inventory level of the maximum number of receivers; and $2)$ to minimise the time a vehicle drives empty performing load balancing operations. Our conjecture is that the faster the fleet gets rebalanced, the higher is the demand served. 

We start by defining the \emph{utility value} $J^k_{ij}$ assigned to the pair of zones $i$ and $j$ as follows:
\begin{equation}
    J^k_{ij}=
     \begin{cases} 
       T_R - T(i,j) & \text{if } i \in \mathcal{F}^k, j \in \mathcal{R}^k \\
       - T_R       & \text{otherwise}
  \end{cases}
  \label{eq:obj_flow}
\end{equation}
where $T(i,j)$ is the travel time between centroids of zones $s_i$ and $s_j$ (see Section~\ref{sec:system}). Owing to Eq.~\eqref{eq:obj_flow}, the utility of a feeder-receiver pair decreases as the time to complete a single relocation increases. The optimisation problem we formulate based on~\eqref{eq:obj_flow} is described in Eqs.~\eqref{eq:flux_obj}-\eqref{eq:flux_set}. The decision variable $x^k_{ij}$ expresses the number of vehicles that should be relocated from feeder $i$ to receiver $j$.
\begin{Problem}
\caption{: Optimal selection of relocation flows}
\begin{align}
    \max &  \sum_{i \in \mathcal{S}} { \sum_{j \in \mathcal{S}} { J^k_{ij} x^k_{ij}}} \label{eq:flux_obj}\\
    s.t. 
    & \sum_j {x^k_{ij}} \leq b_i^k  \quad \forall i \in \mathcal{F}^k \label{eq:flux_F}\\
    & \sum_i {x^k_{ij}} \leq -b_j^k     \quad \forall j \in \mathcal{R}^k \label{eq:flux_R}\\
    & x^k_{ij} \in \mathbb{N}_0  \label{eq:flux_set}
\end{align}
\label{pr:fluxes}
\end{Problem}

The objective function~\eqref{eq:flux_obj} maximises the overall utility of the relocation process. Since each relocated vehicle satisfies one user request, we assign the same utility to each relocated vehicle. Implicitly, this also means that every successful trip returns the same profit to the car-sharing operator. Constraint~\eqref{eq:flux_F} ensures that the number of vehicles relocated from feeder $i$ cannot exceed the positive imbalance at the station. Similarly, constraint~\eqref{eq:flux_R} ensures that the number of vehicles relocated to receiver $j$ does not exceed the negative imbalance at the zone $j$. Constraint~\eqref{eq:flux_set} ensures that $x^k_{ij}$ belongs to the set of natural numbers (including 0). Note that Problem~\ref{pr:fluxes} is a variation of a 0-1 Multiple Knapsack Problem (MKP), which is an NP-hard problem. However, several relaxation techniques and heuristic approaches exist to obtain tight upper bounds of the problem solution in polynomial time~\cite{burkard_assignment_2009}. 
\vspace{-0.2cm}
%
%---------------------------------------------------
\subsection{Schedule of relocation tasks\label{sec:relocation_task}}
\noindent
The output of the second stage of the decision chain of Fig.~\ref{fig:decision_chain} is a list of matched feeder-receiver pairs with associated the total number of vehicles to relocate (i.e. a relocation flow). Now, we want to answer the following research question: which is the most \emph{efficient} way to relocate $x^k_{ij}$ vehicles from station $i$ to station $j$. Clearly, the answer depends on the constraints imposed by the specific redistribution technique that is employed. In the following sections, we develop a range of optimisation models for operator-based, user-based and robotic relocation schemes. The key novelty of our solutions is to consider stackable cars that can be driven in a train. Our optimisation framework is general enough to be applied also to car-sharing systems using conventional cars. 
%!!!COMMENT!!!
%\textbf{necessita di piu' spiegazione: eta=1 non vuol dire necessariamente che c'e' un treno di veicoli lungo 2, può anche essere concettualmente equivalente a relocation con due operatori e service car.}
% in realtà, eta=2, perchè uno dei veicoli è quello del relocator. 
% Sono d'accordo. Ho eliminato da qui il parametro eta che non veniva spiegato bene ed ho messo qualche dettaglio in più dopo prima della formula (14) 
%
%
%---------------------------------------------------
\subsubsection{Operator-based scheme\label{sec:relocation_operator_based}}
Let $\mathcal{O}=\{o_1,o_2, \ldots, o_M\}$ denote the set of $M$ professional drivers (\emph{relocators}) that are employed by the car-sharing operator to relocate vehicles. Without ambiguity, in the following, we use the index $u$ to refer to relocator $o_u$.  % uso di "operator" con due significati diversi potrebbe creare confusione
\begin{figure}[tbhp]
    \centering
    \includegraphics[trim={3cm 1.2cm 8cm 6.5cm},clip,angle=0,width=0.7\columnwidth]{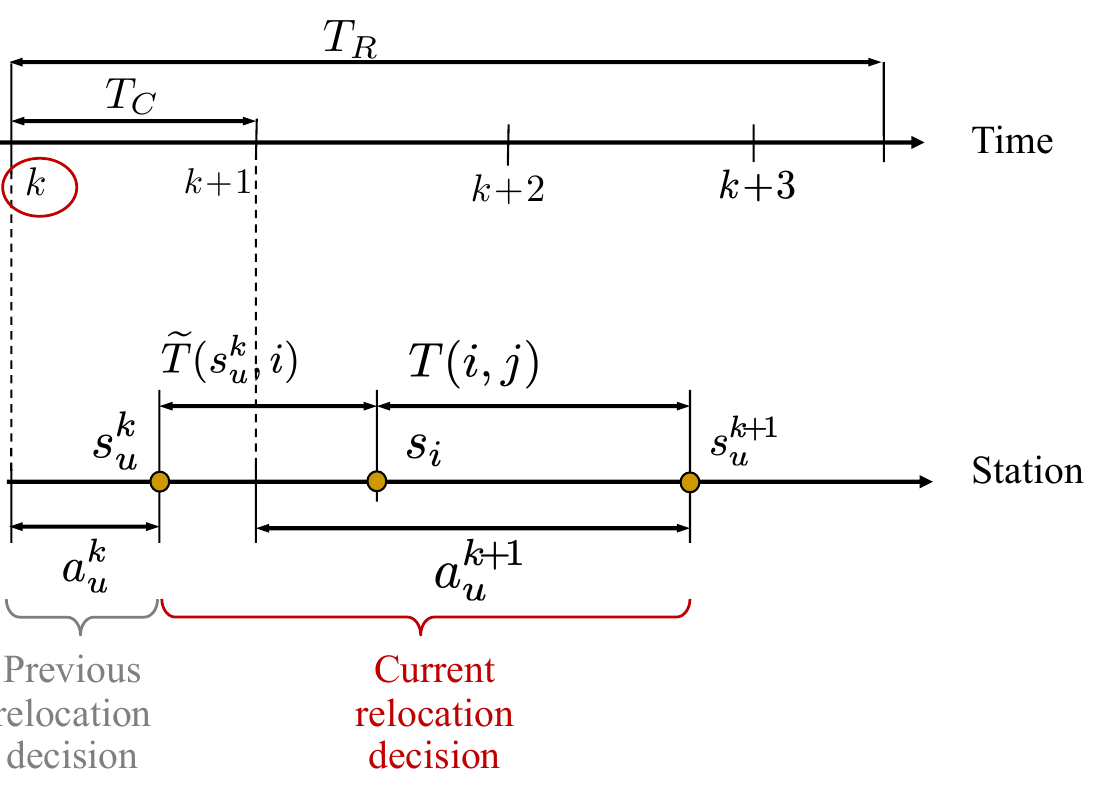}
    \caption{Example of relocation decisions for operator $o_u$ at decision interval $k$ in the case $y^k_{uijl}=1$.}
    \label{fig:scheme}
    \vspace{-0.3cm}
\end{figure}

Let us consider the $k$-th decision interval. To account for varying working shifts, let $\mathcal{\widehat{O}}^k \subseteq \mathcal{O}$ be the subset of relocators that are available during the $k$-th decision interval. Let us assume that relocator $u \in \mathcal{\widehat{O}}^k$ is already relocating a vehicle at time $kT_C$. Then, two variables can be associated with relocator $u$: $1)$ $s_u^k$, defined as the destination zone of $u$, and $2)$ $a_u^k$, defined as the residual time (in time slots) to reach $s_u^k$ (see Fig.~\ref{fig:scheme}). A relocation flow can be split into multiple relocation tasks depending on the maximum allowable train size. Specifically, we assume that each relocator uses a service car to move between zones, and that up to $(\eta+1)$ vehicles can be stacked together. Thus, each operator can relocate at most $\eta$ customer vehicles during a single relocation task, since one vehicle in the relocated train is the service used by the operator to travel to the next feeder. Hence, a relocation flow $x^k_{ij}$ can be split into $L^k_{ij}=1+\floor{x^k_{ij}/(\eta)}$ relocation tasks, each one used to relocate $p^k_{ijl}$ vehicles. In principle, a conventional car-sharing system can be modelled by setting $\eta=1$ and assuming that the service car is driven by a second operator.

By definition, it holds that:
\begin{align}
    x_{ij}^k &=\sum_{l=1}^{L_{ij}^k} p_{ijl}^k\\
    p_{ijl}^k &\leq \eta \; .
\end{align}
For the sake of notation brevity, we introduce the set $\mathcal{P}_{ij}^k=\{p^k_{ij1},p^k_{ij2},\ldots,p^k_{ijL^k_{ij}}\}$ that contains the relocation tasks that compose the relocation flow $x^k_{ij}$. The optimisation problem we formulate to assign relocators to relocation tasks is described in Eqs.~\eqref{eq:assigobj}-\eqref{eq:assignment_limits}. The decision variable $y^k_{uijl}$ expresses the assignment of relocator $u$ to relocation task $l$ in the set $\mathcal{P}^k_{ij}$. 
\begin{Problem}[ht]
\caption{: Optimal assignment of relocation tasks to operators}
\begin{align}
    \max & \sum_{u \in \mathcal{\widehat{O}}^k}  \sum_{i \in \mathcal{F}^k}\sum_{j \in \mathcal{R}^k} \sum_{l \in \mathcal{P}_{ij}^k}  { \Big ( p_{ijl}^k - \frac{ a_u^k + \widetilde{T} (s_u^k,i) } {T_R} \Big ) y^k_{uijl}  } \label{eq:assigobj}\\
    s.t. 
    &  \sum_{i \in \mathcal{F}^k}\sum_{j \in \mathcal{R}^k} \sum_{l \in \mathcal{P}_{ij}^k}  y^k_{uijl} \leq 1  \quad \forall u \in \mathcal{\widehat{O}}^k \label{eq:assignment_onetask}\\ % un utente può fare solo un'operazione
    & \sum_{u \in \mathcal{\widehat{O}}^k}  {y^k_{uijl}}  \leq 1 \qquad  \forall i \in \mathcal{F}^k, \forall j \in \mathcal{R}^k, \forall l \in \mathcal{P}_{ij}^k \label{eq:assignment_oneop} \\ % un'operazione può essere fatta solo da un'utente alla volta 
    & y^k_{uijl} \Big (a_u^k + \widetilde{T}(s_u^k,i) + T(i,j) \Big ) \leq T_R \label{eq:assignment_timelimit} \\ \nonumber
    & \qquad \qquad \qquad \forall u \in \mathcal{\widehat{O}}^k, \forall i \in \mathcal{F}^k, \forall j \in \mathcal{R}^k, \forall l \in \mathcal{P}_{ij}^k  \\
    & y^k_{uijl} \in \{0,1\} \label{eq:assignment_limits}
\end{align}
\label{pr:assignment}
\end{Problem}

The objective function in Eq.~\eqref{eq:assigobj} expresses the maximisation of the relocation efficiency, accounting not only for the vehicle relocation but also for relocator rebalancing. Specifically, the first part of Eq.~\eqref{eq:assigobj} ensures that the maximum number of vehicles are relocated. The second part of Eq.~\eqref{eq:assigobj} refers to the total time that relocator $u$ spends in completing the ongoing relocation tasks (i.e. $a^k_u$) and reaching the feeder zone $i$ from the destination zone $s^k_u$ of the previous relocation task. The former time is denoted with $\widetilde{T} (s_u^k,i)$, and, in theory, it can be different from $T(s_u^k,i)$ (for instance, because the relocator travels between stations using a bike). Hence, this second term minimises the delay incurred by an operator before starting a new relocation task. Given that the second term takes values in the range $[0,1]$, our formulation prioritises over the number of relocation trips. Constraint~\eqref{eq:assignment_onetask} restricts to one the number of new tasks that a relocator can perform in $T_R$. Note that the model can be expanded to remove this restriction but in real-world cases, the length of the $T_R$ period is comparable to the $T(i,j)$ values and do not allow the same relocator to perform multiple relocation tasks within the same relocation period. Constraint~\eqref{eq:assignment_oneop} ensures that a relocation task is assigned to a single relocator. Constraint~\eqref{eq:assignment_timelimit} restricts the assignment of a relocation task to relocators that can complete it within the relocation interval.

After assigning relocators to relocation tasks we can update the $s^{k \!+\!1}_u$ and $a^{k \!+\!1}_u$ variables as follows:
\begin{align}
a_u^{k \!+\!1} & =\begin{cases} \max ( 0, a^k_u + \Big [  \widetilde{T}(s_u^k,i) \!+\! T(i,j) \Big ]  \!-\! T_C ) & \textrm{if } y_{uijl}^k =1 \label{eq:update_a}\\
\max (0 ,  a^k_u-T_C) & \textrm{otherwise} 
\end{cases} \\
s_u^{k \!+\!1} & =\begin{cases} j & \textrm{if } y_{uijl}^k =1 \label{eq:update_s}\\
s_u^k & \textrm{otherwise} 
\end{cases}
\end{align} 
It is easy to observe that when relocator $u$ is not performing a relocation task in the $k$-th decision interval, then $a_u^{k \!+\!1}$ is the previous residual time minus the length of the decision interval (the residual time cannot be less than zero). If a relocation task is assigned, this is added to the next residual. The new destination station $s^{k \!+\!1}_u$ for relocator $u$ is $j$ if $y^k_{uijl}=1$, otherwise it remains unchanged.
%
%
%---------------------------------------------------
\subsubsection{User-based scheme\label{sec:relocation_user_based}}
\noindent
In traditional user-based relocation schemes, the car sharing operator needs to convince customers, leveraging fare discounts, to change the destination of their trips~\cite{febbraro_user_based_2018}. Stackable vehicles add a new dimension to the problem: customers can now take an additional vehicle with them, and this can be beneficial for the rebalancing, even without a change in the destination (e.g., if the customer is already travelling towards a zone that needs additional cars). Note that we assume that only professional drivers with a bus driving license can relocate trains of more than 2 vehicles, while users with conventional driving license can take at most an additional car. Furthermore, customers are not required to manually tow vehicles by themselves as vehicle towing occurs at dedicated stations in a semi-automated way\footnote{An example of automatic towing operations of real ESPRIT vehicles is available at \url{https://youtu.be/ayZ4-7O6rSs}.}. Obviously, intervening only on the number of towed cars and not on the destination may be ineffective in some situations (e.g.,  there might not be enough users travelling in the ``right'' direction for relocation), but it still provides an improvement over traditional approaches with very limited inconvenience for the customers. Thus, in the following we focus on this scenario, and we leave for future work the investigation of the willingness of customers to change their destinations as a function of the offered fare discount. In the following, we simply assume that a user accepts the request from the car-sharing of driving a train of two vehicles with a constant probability equal to $\gamma$, independently of the price discount offered by the car-sharing operator to reward customers for the relocation of cars. Then, a relocation task is performed from zone $i$ to zone $j$ during the $k$-th decision interval if the following conditions are satisfied: $1)$ a customer heading toward zone $j$ arrives at zone $i$ during the $k$-th decision interval; $2)$ zone $i$ and zone $j$ have a surplus and deficiency of vehicles, respectively\footnote{This condition requires that zone $i$ is a feeder and zone $j$ is a receiver. After each relocation the value of the inventory imbalance at the feeder (receiver) is decreased (increased) by one.};
and $3)$ the customer is willing to relocate a vehicle. The design of optimised pricing schemes for incentivising user-based relocation is left as future work.
%
%
%---------------------------------------------------
\subsubsection{Robotic scheme\label{sec:relocation_robotic}} 
\noindent
It is widely recognised that vehicles having self-driving capabilities enable a more efficient relocation process because empty vehicles can autonomously relocate to a zone when needed without waiting for a relocator or a user. Thus, relocation tasks are only constrained by the availability of cars. 

As explained in Section~\ref{sec:relocation_flows}, the output of the second decision stage is the number $x_{ij}^k$ of vehicles that should be relocated from station $i$ to station $j$. The simplest approach for a robotic relocation scheme would be to transfer vehicles from station $i$ to station $j$ as soon as they are available, and until the number of relocated vehicles is equal to $x_{ij}^k$. However, in our relocation model, we follow the same approach as in~\cite{pavone_robotic_2012}, which assumes that vehicles autonomously relocate themselves from zone $i$ to zone $j$ with a constant rate equal to $\alpha_{ij}$ vehicles. The optimal relocation rate is updated at the beginning of the $k$-th decision interval as follows:
\begin{align}
    \alpha_{ij}^k &= \frac{x^k_{ij}}{T_R- T(i,j)} \; , \label{eq:alpha}
\end{align}
where $x^k_{ij}$ is the solution of Problem~\ref{pr:fluxes}. It is important to point out that we can not use the rebalancing strategy defined in~\cite{pavone_robotic_2012} because users were allowed to wait for an available vehicle, and the deficiency of vehicles was measured in terms of waiting customers. Nevertheless, our model formulation in Problem~\ref{pr:fluxes} and the one in~\cite{pavone_robotic_2012} are equivalent if the surplus of vehicles in the feeders is sufficient to rebalance all receivers. 

%
%---------------------------------------------------
\section{Evaluation\label{sec:evaluation}}
\noindent
To evaluate the effectiveness of the proposed relocation framework, we have implemented the optimisation models described in Section~\ref{sec:decision} in Matlab. We have also developed a discrete-time simulation model of a \emph{station-based} one-way car-sharing system to validate the feasibility of the solutions generated by the optimisation models in realistic settings. Before presenting the results, we describe the simulation setup, the data we use to characterise the demand for pickup and drop-off of vehicles, and the benchmarks we adopt for performance comparison.
%
%
%---------------------------------------------------
\subsection{Data and simulation setup\label{sec:data}}
\noindent
There are a few publicly available data sets about real-world car-sharing services~\cite{boldrini_data_mining_2019}. However, these data sets primarily contain data about pickup and drop-off times/locations of rented vehicles, but they do not disclose trip trajectories or rejected rental requests, as this is private and valuable commercial information. Furthermore, commercial car-sharing systems typically implement only overnight relocation. Hence, daytime demand patterns extracted from these data sets are necessarily balanced (since customers cannot pick up vehicles that are not there), and they are ill-suited for studying the efficiency of relocation policies. To circumvent this limitation, we use the data set of trips by New York's yellow taxis, available from the New York Taxi and Limousine Commission (TLC)~ \cite{city_of_new_york_tlc_nodate}. These trips are likely to be unbalanced and to reflect the effective passenger demand, as taxi drivers can relocate their vehicles without passengers. An extensive analysis of the traffic patterns and demand imbalance properties of this data set are reported in Appendix~B of the supplemental material. We have chosen the first 10 Wednesdays of 2018 (3 January to 7 March) as representative weekdays for our simulations. The number of trips in the data set for each day considered are reported in Table~\ref{tab:trips}. Each day includes over 200,000 trips. We found out that the beyond the classical morning peak period between 8:00AM to 10:00AM (with 300 trip requests per minute), there is also a similar demand peak in the evening between 6:00PM to 8:00PM. A substantial demand is also observed during off-peak periods, with an average of 200 trip requests per minute. 
\begin{table}[t]
\centering
\caption{Number of Trips for the Ten Days Used in the Evaluation (New York City, Taxi Data, Year: 2018).}
\bgroup
\begin{tabular}{l|c|c|c|c|c}
\hline
day & 3 Jan  & 10 Jan & 17 Jan & 24 Jan & 31 Jan \\
\hline
trips & 224062 & 245844 & 261854 & 270451 & 273514 \\
\hline
\hline
day & 7 Feb  & 14 Feb & 21 Feb & 28 Feb & 7 Mar \\
\hline
trips & 273723 & 275086 & 251767 & 264750 & 204263 \\
\hline
\end{tabular}
\egroup
\label{tab:trips}
\vspace{-0.5cm}
\end{table}
Regarding the simulation model, it is a time-stepping simulation with $\tau=1$~minute. At each time step, the simulation model begins by first checking if there is a trip request and if there is a vehicle available. Vehicle are assigned on a first-come-first-served basis. Customers are not allowed to wait at stations for available vehicles. If there are no rental requests, relocation tasks are executed depending on the current status of the system. For simplicity, we assume that a station is deployed at each centroid of a zone. Moreover, trips start from and arrive at the station since we are ignoring access walking times. Travelling times between pairs of zone centroids are assumed constant during the day, and they are estimated as the average duration of the taxi trips in the data set with pickups and drop-offs in those zones. In the following tests, we consider varying fleet size. We run the model in \cite{boyaci_optimization_2015} without relocation to determine the initial position of vehicles at stations than minimise the probability of rejecting trip requests. 
% <-- lasciare queste info anche senza grafico?
%Regarding the operator-based relocation techniques, we assume that a relocator travels from a receiver to a feeder using a bike if vehicles are conventional. In this case, we assume that $\widetilde{T}(i,j) \!=\! 2 \times T(i,j)$. 
% -->

Five different relocation policies are tested in the following experiments: $i)$ the operator-based scheme described in section~\ref{sec:relocation_operator_based} with three different train sizes, $\eta \!=\! 7$ (labelled \textsc{OpR-E7}), $\eta \!=\! 2$ (labelled \textsc{OpR-E2}), and $\eta \!=\! 1$ (labelled \textsc{OpR-E1}); $ii)$ the user-based relocation described in section~\ref{sec:relocation_user_based} (labelled \textsc{UsR}) under the assumption that $\gamma \!=\! 1$ (i.e., customers always accept the relocation offer); and $iii)$ the autonomous relocation scheme described in section~\ref{sec:relocation_robotic} (labelled \textsc{AR}). We recall that with stackable cars the relocator can use a service vehicle that is connected to the train of vehicles to be relocated. The advantage is that the journey from a receiver to a feeder is faster than using alternative transportation like a bicycle or public transport (namely $\widetilde{T}(i,j) \!=\! T(i,j)$). The downside is that in a train of $k$ vehicles, $(k-1)$ vehicles satisfy the relocation needs while one is used by the operator to relocate himself. Note that in \textsc{OpR-E1} the driver is allowed to relocate only one vehicle in addition to his service car. Thus, it is the policy that the most closely resembles an operator-based scheme used in conventional car-sharing systems. If not otherwise stated, the following results are obtained by using the worst-case estimate of the inventory imbalance described in Section~\ref{sec:worst-case_estimate}. The parameter $C^k_i(t)$ in equation~(\ref{eq:worst-case_imbalance}) is estimated using a simple averaging of trip requests for the NYC taxi dataset.

The key performance metric we use to assess system performance is the percentage of rejected trip requests. The trade-offs of fleet size and relocation resources are analysed. From the operator's perspective, we also consider the running time of relocation models and the relocation efficiency, measured in terms of the number of relocation tasks and travelled distances of relocated vehicles.  To compute performance statistics and 95\% confidence intervals, we have replicated each simulation ten times using data from the ten selected days. 
%
%
%---------------------------------------------------
\subsection{Benchmarks\label{sec:benchmarks}}
\noindent
For performance comparison, we consider three representative state-of-the-art approaches. 

The first benchmark is inherited from the body of work about vehicle relocation in bike-sharing systems. Specifically, we have implemented the truck-based relocation algorithm presented in~\cite{caggiani_modeling_2018} (labelled as \textsc{TrR}). We assume that the relocation is carried out by multiple trucks, each one with a capacity of 20 vehicles. The depot from which the trucks start and complete their routes is located in the centre of the operational area. We recall that \textsc{TrR} allows a relocator to visit and drop-off vehicles at multiple stations during the route. This is different from our operator-based relocation method, which does not allow fractional relocation. We recognised that this is an idealised system, as truck-based relocation is not possible with cars due to the difficulty of loading and unloading cars from stations. Nevertheless, a car-sharing system using stackable cars and a bike-sharing system have a somehow similar mobility concept. Thus, the model in~\cite{caggiani_modeling_2018} can provide a useful baseline of the potential benefits of vehicle towing during relocation operations.

The second benchmark is inspired by the operator-based relocation policy developed in~\cite{boyaci_optimization_2015}. As discussed in Section~\ref{sec:relatedwork}, the original model in~\cite{boyaci_optimization_2015} is quite sophisticated as it jointly considers strategic and operational decisions to maximise the net revenue of the car sharing operator. In that study, the number of relocated trips between each pair of origin-destination stations are selected so that the total time spent to relocate vehicles does not exceed the total available working hours for a shift of the relocation personnel. Then, this benchmark relocation scheme, labelled as \textsc{AggR}, minimises the time needed to perform planned relocations under a similar constraint on the maximum total relocation time. As in~\cite{boyaci_optimization_2015}, we assume that demand is known for the whole day, and that an operating day is divided into time intervals (not necessarily equally long) and each operation (i.e. rental, relocation, charging) starts at the beginning and finishes at the end of a time interval\footnote{In the following tests we use 50 time intervals, chosen so that each interval has about the same number of trip requests.}. Moreover, we do not consider the time that the relocation personnel spend moving from a receiver station to a feeder station: we assume that relocators are ``teleported'' to the station where they are needed when their previous relocation trip is finished. Note that it may happen that a relocation trip is planned to start when the vehicle is not yet available. This implies that not all the selected relocation decisions may be feasible, as this depends on the real availability of vehicles and relocators at feeders. Unfeasible relocation plans are simply ignored during the simulations of the real system.

The third benchmark, labelled as \textsc{UB}, consists in the reference model developed in~\cite{repoux_dynamic_2019} (Appendix A in that work), which simultaneously decides which requests to accept and which relocations to perform with in-advance full knowledge of the demand. This model provides an approximate upper bound \added{for conventional car-sharing systems not using stackable cars}, which maximises the number of accepted requests, subject to flow conservation constraints of vehicles and relocators, and vehicle reservation constraints. Note that this model can reject some trip requests even if vehicles are available at the booking times, if this increases the total number of accepted trips. As assumed in~\cite{repoux_dynamic_2019}, the day is divided in periods of 5 minutes, and the starting and ending times of each trip are discretised using the same time granularity. Since the model in~\cite{repoux_dynamic_2019} cannot be solved efficiently for our problem size, we adopt the same approach as in~\cite{boyaci_optimization_2015}, and we assume that relocated vehicles are firstly accumulated at an imaginary hub and then distributed from that hub to the stations. This transformation significantly reduces the number of relocation variables in the model~\cite{repoux_dynamic_2019} and has a limited impact on the results. To reach feasible solutions at our scales, we also relax the integer constraints and we solve the equivalent linear optimisation. We think this is a reasonable approximation given the size of our problem. We note that in general this leads to slightly better results for this model thanks to the reduced constraints.
%
%
%---------------------------------------------------
\subsection{Effect of fleet size\label{sec:fleet}}
\begin{figure}[thbp]
    \centering
        \subfloat[][5000 vehicles]{
    \includegraphics[width=0.8\columnwidth]{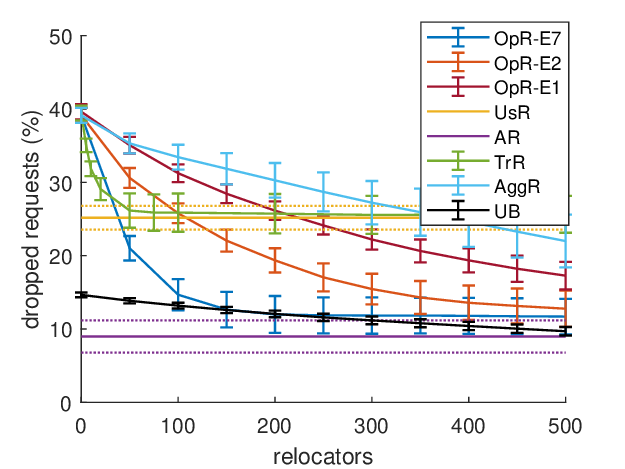}
    \label{fig:dropped_k5000}
} \\%[-2ex]
        \subfloat[][10000 vehicles]{
    \includegraphics[width=0.8\columnwidth]{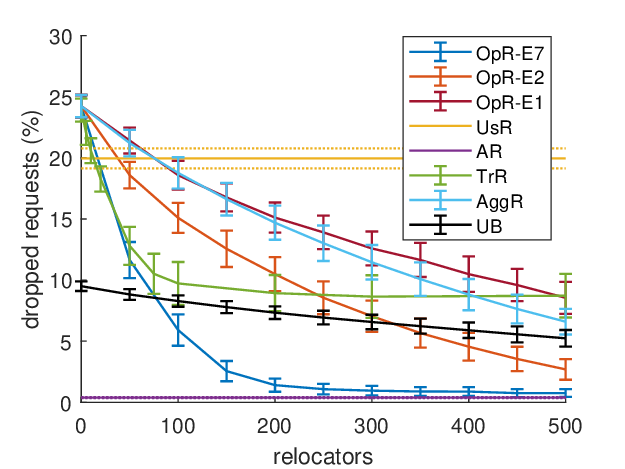}
    \label{fig:dropped_k10000}
}
    \caption{Percentage of rejected trip requests as a function of the number of relocators for (a) 5000 vehicles and (b) 10000 vehicles.} % come separatore delle migliaia meglio usare la virgola (convenzione inglese), abbiamo usato il punto come separatore decimale
    \label{fig:dropped}
\vspace{-0.4cm}
\end{figure}

\noindent
If not otherwise stated, the following results have been obtained using the model parameters listed in Table~\ref{tab:tempoparams}. An extensive parameter-sensitivity analysis has been conducted to select the best model parameters, which is reported in Appendix~C of the supplemental material.
\begin{table}[tbp]
    \centering
    \caption{Best Performing Model Parameters.}
    \begin{tabular}{l|c|c|c}
    Relocation policy    & $T_C$ & $T_R$ & $T_O$ \\
    \hline
        OpR-E7 (worst case)      & 15  & 30  & 45  \\
        OpR-E7 (probabilistic)  & 20  & 40  & 70  \\
        TrR     & 60  & 60  & 120  \\
    \hline
    \end{tabular}
    \label{tab:tempoparams}
    \vspace{-0.4cm}
\end{table}
Fig.~\ref{fig:dropped_k5000} and Fig.~\ref{fig:dropped_k10000} shows the percentage of rejected trip requests versus the number $O$ of operators for the considered relocation policies with a fleet of 5000 and 10000 vehicles, respectively. Clearly, the relative performance of the user-based relocation scheme and the robotic one is invariant with the number of relocators. For the \textsc{TrR} scheme, the number on the $x$ axis represents the number of trucks. 

Important observations can be derived from the results shown in Fig.~\ref{fig:dropped_k5000}. First, the robotic relocation scheme achieves the best performance among the considered strategies, but about 9\% of the trip requests are still rejected. This is not surprising, as the relocation process in the robotic scheme is constrained only by the vehicle availability at the feeders. The user-based relocation scheme produces a 150\% increase in the percentage of rejected trip requests when compared to the robotic scheme. Clearly, the efficiency of the \textsc{UsR} scheme depends on several factors, including the willingness of users to perform relocation tasks, the demand patterns, and the probability that there is a passenger trip from a feeder to a receiver when needed. On the contrary, our \textsc{OpR-E7} scheme approximates the efficiency of the robotic scheme with only 200~relocators (i.e. one relocator every 25 vehicles). Moreover, \textsc{OpR-E7} scheme significantly outperforms the \textsc{AggR} benchmark in all the considered scenarios. Even with a large relocation workforce (i.e. one relocator every 10 vehicles), the number of rejected rental request with \textsc{AggR} is twice as much as that of \textsc{OpR-E7}. Clearly, the efficiency of our operator-based relocation scheme degrades when using smaller train sizes, namely $\eta \!=\! 2$. Nevertheless, \textsc{OpR-E7} and \textsc{OpR-E2} achieve similar performance when the number of relocators is sufficiently large. Interestingly, \textsc{OpR-E1} behaves slightly better than \textsc{AggR} even if both polices relocate a single vehicle. We observe that the car-sharing system is severely underserved when a fleet of 5000 vehicles is used. In these conditions, the \textsc{AggR} model is not able to find a relocation plan that can satisfy the full demand, while \textsc{OpR-E1} only searches for a relocation plan that minimises the time spent relocating vehicles. It is also important to point out that using large trains is not necessarily beneficial for relocation efficiency. Indeed, the performance of the \textsc{TrR} scheme rapidly flattens out after a few trucks are employed. Besides, returning the trucks to a central depot rather than rebalancing them from receivers to feeders introduces an excessive delay in the relocation process. Thus, the truck-based relocation scheme achieves performance that is comparable to the user-based scheme with stackable cars.
The results show that the percentage of dropped trip requests when no relocation is performed is significantly lower with the \textsc{UB} benchmark than the other schemes. This is due to the fact that the \textsc{UB} has a perfect advance knowledge of the trip requests and the model selects the trips to accept so that it minimises the fleet imbalance over the whole day. On the contrary, our system and the other benchmarks assign vehicles to passengers on a first-come first-served basis, which ensures a fair treatment of customers but a less efficient system. In addition, vehicle relocation has a small effect on the performance of the \textsc{UB} benchmark, as trips than cause inventory imbalance are removed in advance by UB. Thus, we can observe that \textsc{OpR-E7} approaches UB performance with 150 relocators and when the number of relocators is sufficiently high (about 500 in the considered scenario), the performance gain of \textsc{UB} model over \textsc{OpR-E2} is limited. 

The results of Fig.~\ref{fig:dropped_k10000} for 10000 vehicles generally confirm the trends observed Fig.~\ref{fig:dropped_k5000}. Clearly, doubling the fleet size increases the capacity of the car-sharing system. Indeed, the performance of all relocation schemes improves, with the robotic scheme that is able to satisfy the demand fully. Interestingly, we can observe that the use of longer road trains ($\eta \!=\! 7$) provides a faster performance gain because there are more opportunities for a relocator to find a full train of vehicles parked at the feeder when starting a new relocation task (see also Table~\ref{tab:trainlength}). Therefore the performance gap between \textsc{OpR-E7} and \textsc{OpR-E2} increases with a fleet of 10000 vehicles. \textsc{AggR} performs slightly  better than \textsc{OpR-E1} with a fleet of 10000 vehicles. In this case, the unfilled demand is low and the relocation policy of \textsc{AggR} is able to determine more efficient relocation plans. As already shown in Fig.~\ref{fig:dropped_k5000}, \textsc{UB} performs significantly better than the other relocation schemes when the number of relocators is low since it can select the most beneficial trips to serve. However, with 10000 vehicles the relative performance gain of UB model is smaller than what observed with 5000 vehicles. In addition the efficiency of the UB model increases very slowly when increasing the size of the relocation personnel. Thus, the relocation with stackable cars can outperform the \textsc{UB} benchmark when the number of relocators is higher than 75 for \textsc{OpR-E7} and 300 for \textsc{OpR-E2}.

\begin{table}[t]
    \centering
    \caption{Spatio-temporal analysis of dropped requests (\%). The results refer to a relocation staff of 200 drivers}
    \begin{tabular}{|l|c|c|c|}
    \hline
    Spatio-temporal domain & No relocation & OpR-E7 & \textsc{AggR}  \\ 
    %domain & relocation &   & \\
    \hline
    \hline
    \multicolumn{4}{|c|}{5000 vehicles}\\
    \hline
    Morning peak (8am-10am)        & 42.48   &  17.55 & 31.71\\
    Off-peak (noon-2pm)    & 27.71   &  4.82 & 18.33\\
    \hline
    5 most imbalanced & 48.56   &  10.82 & 39.86\\
    %5 most negatively imbalanced & 0.15   &  2.93 & 0.32\\
    \hline
    \hline
    \multicolumn{4}{|c|}{10000 vehicles}\\
    \hline
    Morning peak        & 19.57  & 1.18 & 7.62 \\
    Off-peak    & 18.57  & 0.03 & 10.37\\
    \hline
    5 most imbalanced & 34.66 & 2.09 & 23.53\\
    %5 most negatively imbalanced & 0.00 & 2.47 & 0.03\\
    \hline
    \end{tabular}
    \label{tab:spatio-temporal}
    \vspace{-0.4cm}
\end{table}
    %             K=5000
    %   0.42479      0.17553      0.31712
    %   0.27714     0.048189      0.18335
    %   0.48559       0.1082      0.39863
    % 0.0014501     0.029292    0.0031903
    %             K=10000
    %   0.19565      0.01177     0.076242
    %   0.18572   0.00025592      0.10368
    %   0.34657     0.020932      0.23527
    %         0     0.024652   0.00029002
%
The results in Fig.~\ref{fig:dropped_k5000} and Fig.~\ref{fig:dropped_k10000} show the performance gains over the whole day. Table~\ref{tab:spatio-temporal} reports the percentage of rejected trips in particularly relevant time intervals or stations for a scenario with 200~relocators. Specifically, we investigate the relocation efficiency during the morning peak, from 8:00AM to 10:00AM, and the off-peak period, from noon to 2:00PM. Moreover, we analyse the relocation efficiency in the five stations that show the deepest traffic imbalance. The results indicate that the off-peak period benefits more than the morning peak, with a five-fold decrease of dropped requests for 5000 vehicles, but a two-fold decrease of dropped requests is achieved also in the off-peak period. As expected the receiver stations with the largest deficiency of vehicles are the ones with the highest percentage of dropped requests, but they also benefit the most from relocated vehicles. 
%
%
%---------------------------------------------------
\subsection{Relocation efficiency\label{sec:efficiency}}
\begin{figure*}[ht]
    \centering
    \subfloat[][]{
    \includegraphics[width=0.25\textwidth]{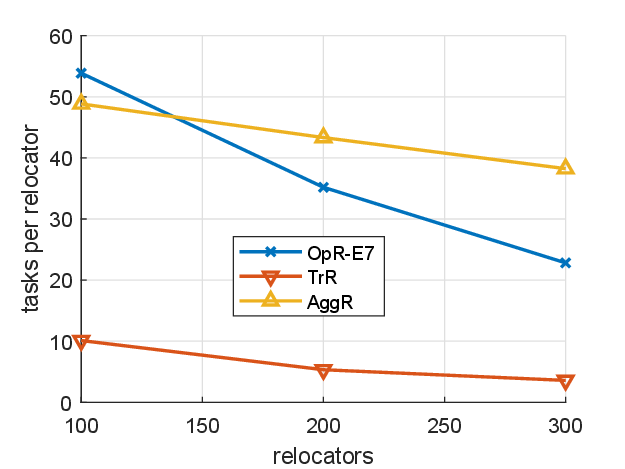}
    \label{fig:opstats1}
    }
    \subfloat[][]{
    \includegraphics[width=0.25\textwidth]{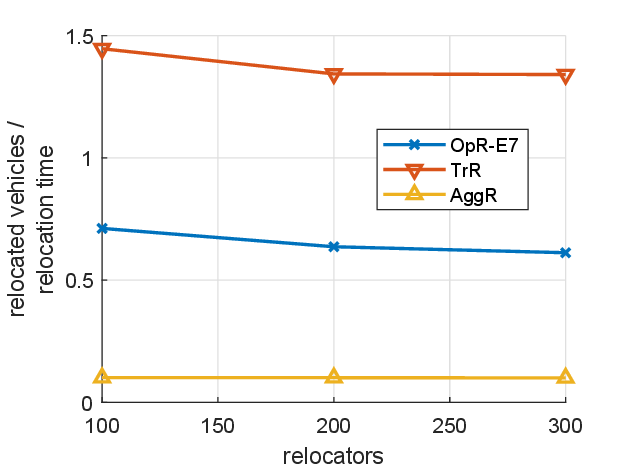}
    \label{fig:opstats2}
    }
    \subfloat[][]{
    \includegraphics[width=0.25\textwidth]{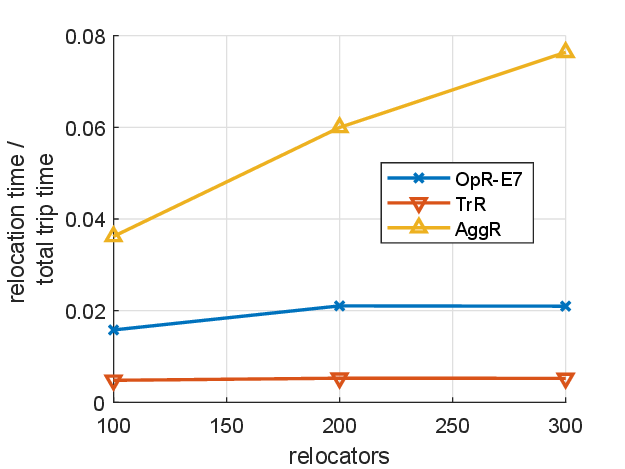}
    \label{fig:opstats3}
    }
    \caption{Relocation efficiency in terms of: a) number of relocation tasks per relocator; b) number of relocated vehicles over daily relocation time; c) daily relocation time over the total duration of trips with passengers. Fleet size equal to 10000 vehicles.
    }
    \vspace{-0.4cm}
\end{figure*}

\noindent
In this section we investigate the efficiency of the operator-based relocation schemes from the car-sharing operator's perspective. Intuitively, it is important for the car-sharing operator to efficiently utilise the relocation workforce by ensuring that each relocator completes the largest possible number of tasks and that the distance vehicles drive empty is minimised. To this end, Fig.~\ref{fig:opstats1} shows the number of tasks each relocator performs on average with a fleet of 10000 vehicles. Our results show that \textsc{AggR} uses relocators more frequently than other schemes. This can be explained by noting that both \textsc{OpR-E7} and \textsc{TrR} rapidly reach their optimal performance (see Fig.~\ref{fig:dropped_k10000}), and the relocation efficiency is mainly constrained by vehicle availability rather than relocator availability. In this condition, the more relocators are employed, the longer they are idle. Clearly, the simultaneous relocation of multiple vehicles with a single relocation task also allows \textsc{OpR-E7} and \textsc{TrR} to relocate vehicles more rapidly than \textsc{AggR} (see Fig.~\ref{fig:opstats2}). Finally, Fig.~\ref{fig:opstats3} shows the ratio between the total time vehicles drive empty for relocation (namely, the relocation time) and the time vehicles drive with passengers. We point out that a train of relocated vehicles contributes to the total relocation time only once. The results show that \textsc{OpR-E7} significantly outperforms \textsc{AggR} as it ensures a higher utilisation of the shared vehicles for profitable trips.   
%
%
%% ADD STATISTICS:
% Average daily travel distance per vehicle (km) 
% Total daily vehicle travel distance (km) 
% Total daily served trip distance (km) 
% Total daily unoccupied travel distance (km) 

To conclude this study, Table~\ref{tab:trainlength} shows statistics about train length for \textsc{OpR-E7} for different fleet sizes and numbers of relocators. The results show that between 60\% and 80\% of relocation actions are performed with the maximum train length of 8 vehicles. However, a non-negligible fraction of relocation trips involves smaller trains, which allows taking advantage of feeder stations with a small surplus of vehicles. The length of vehicle trains increases with the number of vehicles and decreases with the number of available relocators. This is because the larger the fleet, the larger the number of vehicles that are available at each station, and relocators give higher priority to tasks that relocate the most vehicles (see problem formulation in~\ref{pr:assignment}). Finally, Table~\ref{tab:trainlength} shows the fraction of time relocators spend rebalancing themselves when completing a relocation task. The results show that about 50\% of the travel time of operators is used to travel from receiver to feeder stations. This further justifies the \textsc{AggR}, which considers only vehicle relocation and not operator relocation in the formulation of the relocation problem. 
\begin{table}[ht]
    \centering
    \caption{Length of trains for OpR-E7 (\%). $O$ denotes the number of relocators, and $K$ the fleet size.}
    \begin{tabular}{c|c c c c}
    \hline
    & \multicolumn{2}{c}{K=5000} & \multicolumn{2}{c}{K=10000}\\
    Length (x) & O=100 & O=200 & O=100 & O=200  \\
    \hline
    $x< 3 $  & 14.0 & 16.6 & 6.0 & 9.7\\
    $3 \leq x < 5$  & 12.0 & 12.4 & 6.9 & 8.9\\
    $5 \leq x < 8 $  & 9.2 & 10.7 & 6.0 & 7.5\\
    $ x=8$  & 64.8 & 60.4 & 81.1 & 73.9\\
    \hline
    \end{tabular}
    \label{tab:trainlength}
    \vspace{-0.4cm}
\end{table}
\begin{table}[ht]
    \centering
    \caption{Fraction of time that relocators spend travelling from receiver to feeder stations.}
    \begin{tabular}{c|c c c c}
    \hline
    & \multicolumn{2}{c}{K=5000} & \multicolumn{2}{c}{K=10000}\\
    Model & O=100 & O=200 & O=100 & O=200  \\
    \hline
    \textsc{OpR-E2}  & 52.8 & 53.1 & 52.0 & 52.6\\
    \textsc{OpR-E7}  & 52.0 & 50.4 & 52.4 & 51.0\\
    \hline
    \end{tabular}
    \label{tab:drivingmepty}
    \vspace{-0.4cm}
\end{table}
%
%

%
%---------------------------------------------------
\section{Conclusions\label{sec:conclusions}}
\noindent
Most of the research effort on car sharing systems is focused on vehicle relocation, which is considered the most difficult operational aspect of these systems. Previously proposed solutions are generally very complex, slow and with poor scalability. This limits their practical implementation. In this work, we presented a novel fast and scalable relocation optimisation framework considering stackable cars. We use problem decomposition to split the relocation problem into three simpler sub-problems. We showed that our algorithm outperforms previously proposed relocation algorithms while keeping computational complexity low. The capabilities of the proposed approach are demonstrated on a large scale data set comprising over 200,000 taxi trips per day in New York. Furthermore, our approach relies on limited aggregate information about the demand and does not require detailed advance knowledge, which is generally not available in real-world cases.

%As future work, we plan to extend our formulation to deal with electric vehicles and limited driving ranges. Furthermore, vehicle relocation in hybrid car-sharing systems that allow trip reservation in addition to on-demand rentals is also an interesting research problem. Finally, suitable incentive mechanisms for achieving the desired level of user participation in a user-based relocation scheme should also be investigated. 
%
%---------------------------------------------------
\section*{Acknowledgements}
\noindent
This work is partly supported by EIT Climate-KIC under the DRIVE2 (200286) project, and by CNR under the TIRS (FOE 2020) project.

\bibliographystyle{ieeetr}
\bibliography{carsharing}

\end{document}

\appendices
%
%--------------------------------------------------------- 
\section{Pseudocode of the Relocation Algorithms}\label{app:code}
\noindent 
In this Appendix we present a pseudo-code of the algorithm that is implemented in the proposed multi-stage decision support system for vehicle relocation. More precisely, Algorithm~\ref{s-algo:relocation} is executed by the decision tool at the beginning of each decision point $k$. It requires as input the initial state of each zone $i$, and the predicted vehicle arrival and departure processes of that zone; while it provides as output the set of relocation tasks that should be performed. We remind that the high-level description of the functional architecture of the proposed system is provided in Section~\ref{sec:decision}, while the formal description of the proposed relocation solutions is provided in Section~\ref{sec:relocation}. It is also important to point out that the algorithm described in the following has been implemented in Matlab as a stand-alone toolbox for computing relocation plans of a general car-sharing system. 
\begin{algorithm}[]
\renewcommand{\algorithmicrequire}{\textbf{Input:}}
\renewcommand{\algorithmicensure}{\textbf{Output:}}
\caption{Relocation Algorithm}\label{s-algo:relocation}
\begin{algorithmic}[1]
\Require $v^k_i,C^k_i,D^k_i, R^k_i, F^k_i(v^k_i)$
\Ensure Set of relocation tasks to perform during the $k$-th decision interval
\State
\For{$i \in \mathcal{S}$}
\If{(predictor type = \textsc{Worst\_Case})} \label{s-ln:worst}
\State $b^k_i = \min\limits_{t=1,\ldots,n_O} I^k_i (t) $ \label{s-ln:worst_e}
\ElsIf{(predictor type = \textsc{Probabilistic})}
\If{$F^k_i(v^k_i) >0$} 
\State $b^k_i \gets \max \{x \in \mathbb{N}  | F^k_i(v^k_i+x) - \epsilon \le 0 \}$ \label{s-ln:prob_e}
\ElsIf{$F^k_i(v^k_i) = 0$}
\State
\State $b^k_i \gets \min \{ x \in \mathbb{N} | F^k_i(v^k_i-x) - \epsilon \ge 0 \}$
\EndIf
\EndIf
\State 
\If{$b^k_i >0$} \label{s-ln:feed}
\State $\mathcal{F}^k \gets \mathcal{F}^k \bigcup \; \{i\}$
\ElsIf{$b^k_i <0$}
\State $\mathcal{R}^k \gets \mathcal{R}^k \bigcup \; \{i\}$
\EndIf \label{s-ln:rec}
\EndFor
\State 
\State Solve Problem~\ref{pr:fluxes} to compute $x^k_{ij}$   \label{s-ln:prob1}
\State 
\Switch{relocation scheme}

\Case{\textsc{Operator-Based}} \label{s-ln:operator}
\State $\mathcal{P}_{ij}^k=\{p^k_{ijl} \;|\; \sum_{l=1}^{L_{ij}^k} p_{ijl}^k \;,\; p_{ijl}^k \leq \eta\}$
\State Solve Problem~\ref{pr:assignment} to compute $y^k_{uijl}$ ;  \label{s-ln:prob2}
\For{$u \in \mathcal{\widehat{O}}^k$}
\State Update $a_u^{k \!+\!1}$ using Eq.~(\ref{eq:update_a});
\State Update $s_u^{k \!+\!1}$ using Eq.~(\ref{eq:update_s});
\EndFor
\EndCase

\Case{\textsc{User-Based}} \label{s-ln:user}
\State Wait for a trip request $\widehat{r}$ \label{s-ln:wait_trip}
\State $\hat{i} \gets \texttt{Origin}(\widehat{r})$
\State $\hat{j} \gets \texttt{Destination}(\widehat{r})$
\State $\hat{c} \gets \texttt{Customer}(\widehat{r})$
\If{$(\hat{i} \in \mathcal{F}^k \land \hat{j} \in \mathcal{R}^k) \land x^k_{ij}>0 \land \Phi(\hat{c})>\gamma )$}
\State $\hat{c}$ relocates a vehicles;
\State $x^k_{ij} \gets x^k_{ij} -1$
\EndIf
\EndCase

\Case{\textsc{Robotic}} \label{s-ln:robotic}
\For{$i \in \mathcal{F}^k, j \in \mathcal{R}^k$}
\State $\alpha_{ij}^k = \frac{x^k_{ij}}{T_R- T(i,j)}$
\StartT
\State \texttt{\texttt{sleep}($\alpha_{ij}^k T_C$)}
\If{($v^k_i(\texttt{now}) > 0$)}
\State one vehicle relocates to station $j$
\EndIf
\EndT{end of the $k$-th decision interval} \label{s-ln:end}
\EndFor
\State
\EndCase

\EndSwitch
\end{algorithmic}
\end{algorithm}
The input variables of Algorithm~\ref{s-algo:relocation} are the current state of zone $i$ at decision point $k$, expressed in terms of the number $v^k_i$ of available vehicles, and the predictions of the  future demand during the $k$-th decision interval. These input data is used by the algorithm to compute the inventory imbalance depending on the type of estimator the car-sharing operator decides to use, either worst-case (line~\ref{s-ln:worst_e}) or probabilistic (line~\ref{s-ln:prob_e}). Then, the algorithm leverages the inventory imbalance to determine the set of feeder and receiver zones (lines~\ref{s-ln:feed}-\ref{s-ln:rec}). After this, the Problem~\ref{pr:fluxes} is solved to determine the relocation flows that are needed to rebalance the fleet of vehicles (line~\ref{s-ln:prob1}). Afterwards, the relocation algorithms differentiates its operation depending on the relocation strategy adopted by the car-sharing service provider. If an operator-based relocation approach is used (line~\ref{s-ln:operator}, Problem~\ref{pr:assignment} is solved to optimally assign the relocation tasks to the operators. If a user-based relocation approach is used, the algorithm waits for a trip request from customer $\hat{c}$. Then, a relocation task is assigned to that customer only if he/she will travel from a feeder to a receiver zones, the feeder zone has still available vehicles in excess and the customer is willing to participate to the relocation process (i.e. $\Phi(\hat{c})> \gamma$). Finally, if autonomous vehicles are used, the algorithm sets a periodic timer with frequency $\alpha^k_{ij}$ for each pair of feeder and receiver zones. When the timer associated to the pais $\{i,j\}$ expires, if a vehicle is available in  zone $i$, it autonomously relocate to zone $j$. This sequence of relocations continues until the end of the decision interval (line~\ref{s-ln:end}). 
%
%
%--------------------------------------------------------- 
\section{Data and simulation setup}\label{s-app:data}
\begin{figure}[ht]
    \centering
    % \subfloat[][]{ % [trim=left bottom right top, clip]
    % \includegraphics[trim={0cm 2.8cm 0cm 1.7cm}, clip,angle=0,width=0.8\columnwidth]{figures/manhattanmap_horizontal.eps}
    % \label{fig:map}
    % } \\
    \subfloat[][]{ % [trim=left bottom right top, clip]
    \includegraphics[trim={4.5cm 0cm 5cm 0cm}, clip,angle=0,width=0.4\columnwidth]{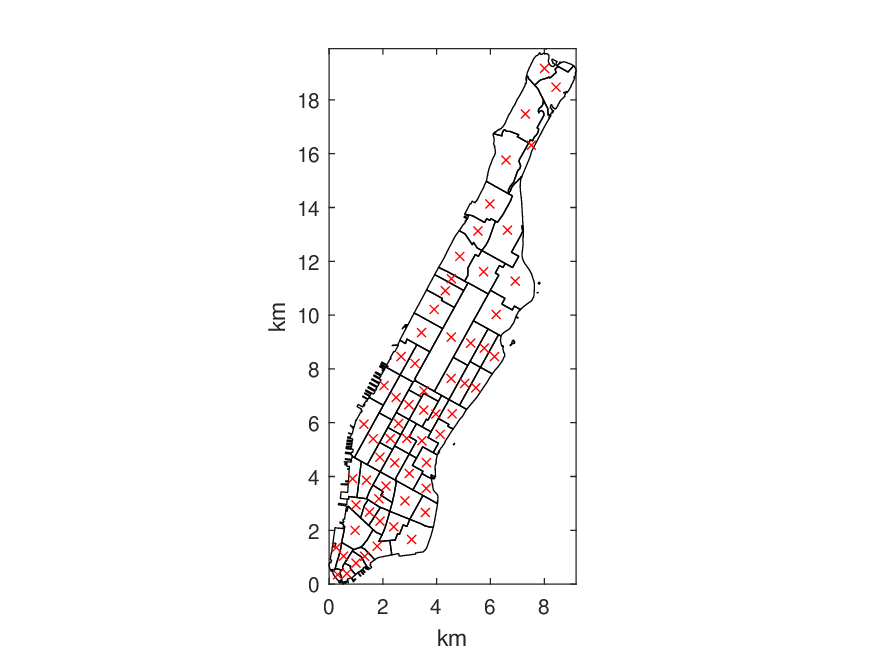}
    \label{s-fig:map}
    }
%    \subfloat[][]{
%    \includegraphics[trim={0cm 0cm 0cm 0cm},clip,angle=0,width=0.8\columnwidth]{figures/requestspermin.eps}
%    \label{fig:arrivals}
%    } \\
    \subfloat[][]{
    \includegraphics[trim={0cm 0cm 0cm 0cm},clip,angle=0,width=0.56\columnwidth]{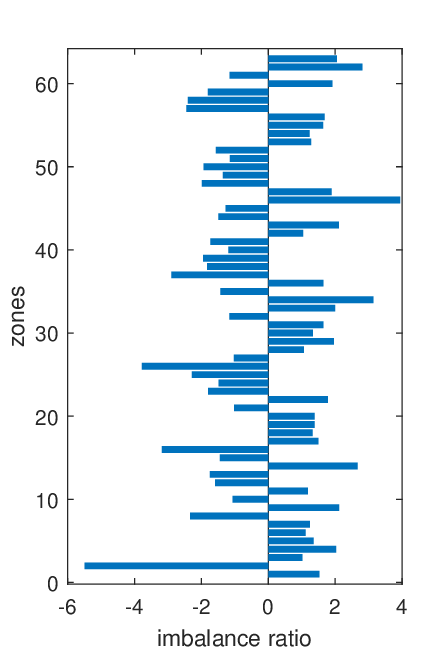}
    \label{s-fig:imbalance}
    }
    \caption{(a) Zoning system used in the data set (crosses are the centroid positions); 
    %b) Temporal distribution of trip frequency on day 9 (28 Feb), whose total number of request is the closest to the median number of request per day in the data set.
    (b) imbalance ratio for each zone during the morning peak period (8am--10am). Positive imbalance indicate more departures than arrivals and vice versa.
    }
\vspace{-0.2cm}
\end{figure}

\noindent 
In this Appendix we extensively discuss the characteristics of the data set we used for the evaluation. Indeed, there are a few publicly available data sets about real-world car-sharing services~\cite{boldrini_data_mining_2019}. However, these data sets primarily contain data about pickup and drop-off times/locations of rented vehicles, but they do not disclose trip trajectories or rejected rental requests, as this is private and valuable commercial information. Furthermore, commercial car-sharing systems typically implement only overnight relocation. Hence, daytime demand patterns extracted from these data sets are necessarily balanced (since customers cannot pick up vehicles that are not there), and they are ill-suited for studying the efficiency of relocation policies. To circumvent this limitation, we use the data set of trips by New York's yellow taxis, available from the New York Taxi and Limousine Commission (TLC)~\cite{city_of_new_york_tlc_nodate}. These trips are likely to be unbalanced and to reflect the effective passenger demand, as taxi drivers can relocate their vehicles without passengers. To better characterise the traffic patterns we introduce the demand imbalance ratio $\rho_i(t_1,t_2)$ of station $i$ during the time interval $[t_1,t_2]$, which is defined as follows:
\begin{equation}
    \rho_i(t_1,t_2) = \begin{cases} {\ \ }\frac{D_i(t_1,t_2)}{O_i(t_1,t_2)} & \textrm{if } D_i(t_1,t_2) \ge O_i(t_1,t_2) \\[0.2cm]
    - \frac{O_i(t_1,t_2)}{D_i(t_1,t_2)} & \textrm{otherwise}
    \end{cases}\; ,
\end{equation}
where $D_i(t_1,t_2)$ ($O_i(t_1,t_2)$) is the number of trips whose destination (origin) is station $i$ during the time interval $[t_1,t_2]$. Intuitively, $\rho_i(t_1,t_2) >> 1$ means that station $i$ accumulates a significant surplus of vehicles, while $\rho_i(t_1,t_2) << -1$ implies that station $i$ has a serious deficiency of vehicles. Now, let us investigate more in depth the traffic characteristics of the NYC demand. To reduce the size of the problem, we consider only trips which have origin or destination in the main island of Manhattan, which is divided into 63 zones, as shown in Fig.~\ref{s-fig:map}. The zoning system is defined by the TLC. Each taxi zone approximates a neighbourhood within a city borough. We have chosen the first 10 Wednesdays of 2018 (3 January to 7 March) as representative weekdays for our simulations. 

Fig.~\ref{s-fig:imbalance} shows the traffic imbalance ratio for the morning peak. It is evident that there are critical service zones where there is up to four times more vehicle departures than vehicle arrivals. These are the traffic ``hot spots'' where relocation is most needed.  
Regarding the simulation model, it is a time-stepping simulation with $\tau=1$~minute. At each time step, the simulation model begins by first checking if there is a trip request and if there is a vehicle available. Vehicle are assigned on a first-come-first-served basis. Customers are not allowed to wait at stations for available vehicles. If there are no rental requests, relocation tasks are executed depending on the current status of the system. For simplicity, we assume that a station is deployed at each centroid of a zone. Moreover, trips start from and arrive at the station since we are ignoring access walking times. Travelling times between pairs of zone centroids in Fig.~\ref{s-fig:map} are assumed constant during the day, and they are estimated as the average duration of the taxi trips in the data set with pickups and drop-offs in those zones. In the following tests, we consider varying fleet size. We run the model in \cite{boyaci_optimization_2015} without relocation to determine the initial position of vehicles at stations than minimise the probability of rejecting trip requests. 
% <-- lasciare queste info anche senza grafico?
%Regarding the operator-based relocation techniques, we assume that a relocator travels from a receiver to a feeder using a bike if vehicles are conventional. In this case, we assume that $\widetilde{T}(i,j) \!=\! 2 \times T(i,j)$. 
% -->
%
%
%
%
%--------------------------------------------------------- 
\section{Supplemental Results}\label{s-app:results}
\noindent 
In this Appendix we report additional results that are helpful to demonstrate the effectiveness of the proposed relocation methods.
%
%
%---------------------------------------------------
\subsection{Sensitivity analysis for parameter selection\label{s-sec:parameter}}
\noindent
\begin{figure*}[ht]
    \centering
    \subfloat[][100 relocators]{
    \includegraphics[width=0.3\textwidth]{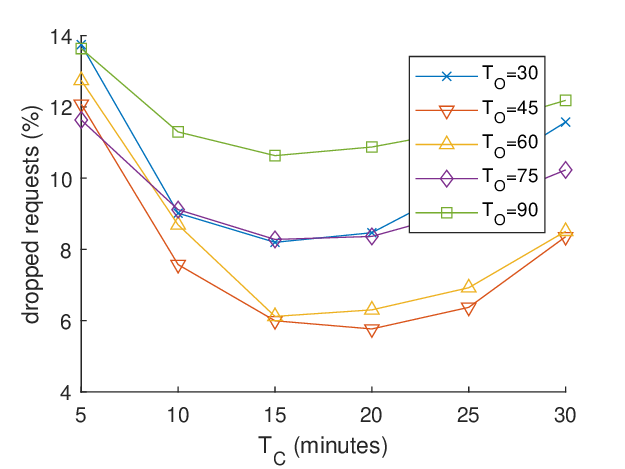}
    \label{s-fig:sensitivity1}
    }
    \subfloat[][200 relocators]{
    \includegraphics[width=0.3\textwidth]{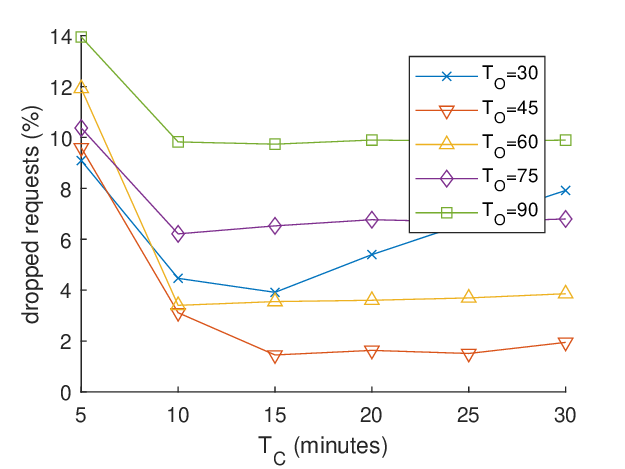}
    \label{s-fig:sensitivity2}
    }
    \subfloat[][300 relocators]{
    \includegraphics[width=0.3\textwidth]{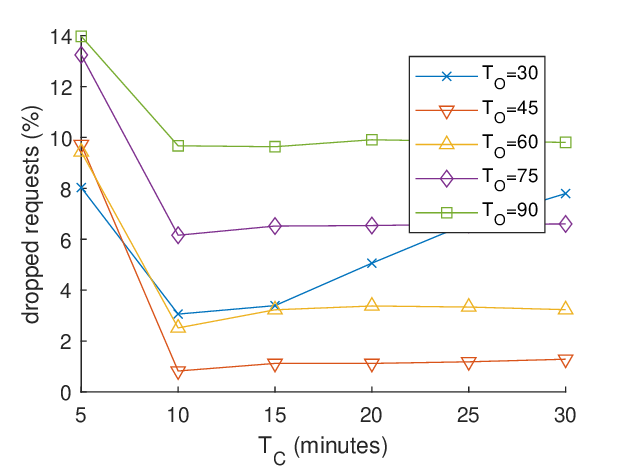}
    \label{s-fig:sensitivity3}
    }
    \caption{\textsc{OpR-E7} scheme: sensitivity analysis of the percentage of rejected user requests versus the time parameters for a fleet of 10000 vehicles and $T_R=30$ for a) 100 relocators; b) 200 relocators; c) 300 relocators. All simulations refer to day 9.}
    \label{s-fig:sensitivity}
\end{figure*}
\begin{table}[tbp]
    \centering
    \caption{Best Performing Model Parameters.}
    \begin{tabular}{l|c|c|c}
    Relocation policy    & $T_C$ & $T_R$ & $T_O$ \\
    \hline
        OpR-E7 (worst case)      & 15  & 30  & 45  \\
        OpR-E7 (probabilistic)  & 20  & 40  & 70  \\
        TrR     & 60  & 60  & 120  \\
    \hline
    \end{tabular}
    \label{s-tab:tempoparams}
    \vspace{-0.2cm}
\end{table}
As explained in Section~\ref{sec:relocation}, the proposed relocation models depend only on three time parameters: $i)$ the planning period $T_C$, $ii)$ the bound $T_R$ on the time that is needed to complete the relocation tasks that is scheduled between a pair of feeder/receiver stations, and $iii)$ the observation period $T_O$. To tune these model parameters we take advantage of the low computation complexity of our model (see Section~\ref{s-sec:cputime} for a more detailed discussion) and the limited number of parameters, and we use a ``brute force'' trial-and-error approach, i.e., a finite number of candidate values is tested over one representative day (namely day 9 in Table~\ref{tab:trips}) to estimate the nearly-optimal setting. Fig.~\ref{s-fig:sensitivity} shows the results of our sensitivity analysis for the \textsc{OpR-E7} scheme for $T_R \!=\! 30$~minutes\footnote{In our data set, 75\% of the trips last less than 30~minutes.}, using a fleet of 10000~vehicles and a varying number of relocators. %Interestingly, we can observe that 
The best set of model parameters is scarcely dependent on the number of relocators, and Table~\ref{s-tab:tempoparams} summarises the best model parameters that we use in the following tests. It is important to point out that the best model parameters also depend on the approach used to estimate the inventory imbalance. We discovered that the probabilistic estimate (see Section~\ref{sec:probabilistic_estimate}) prefers longer control and prediction periods than the worst-case estimate (see Section~\ref{sec:worst-case_estimate}). This can be explained observing that the probabilistic estimate assumes more detailed information on the demand patterns. Thus, a longer control horizon allows the model to better exploit this information, while the worst-case estimate uses only aggregated statistics, whose reliability decreases over time. 

To conclude this section, we observe that several methods can be used to allow a faster tuning of nearly-optimal model parameters. For instance, the concurrent estimation technique can be used to estimate the effect of alternate parameter settings from a single sample path~\cite{cassandras_discrete_2009}. Alternatively, metaheuristic strategies, such as genetic algorithms, can be used.  
%
%
%---------------------------------------------------
%
\subsection{Running time of the optimisation models\label{s-sec:cputime}}
\begin{figure}[tbhp]
    \centering
    \includegraphics[width=0.8\columnwidth]{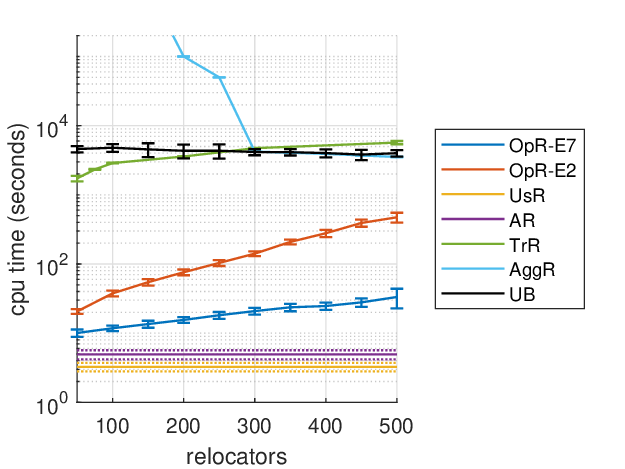}
    \caption{Running time of the optimisation models for a fleet of 10000 vehicles using a log-lin plot (gap tolerance of the optimal value equal to $10^{-2}$).}
    \label{s-fig:cputime1}
\vspace{-0.2cm}
\end{figure}

\noindent
One of the advantages of our multi-stage optimisation is the scalability and low computational complexity, thanks to the problem decomposition. To validate this claim, Fig.~\ref{s-fig:cputime1} shows the running time of the different relocation algorithms for a fleet of 10000 vehicles. 
The models were solved using Matlab intlinprog solver on a server with an Intel~Xeon~E5\nobreakdash-2640~@~2.4~GHz CPU and 16~GB RAM. As expected, the robotic and the user-based relocation schemes require a negligible running time as the scheduling of the relocation tasks does not require solving an integer program. The running times of \textsc{OpR} increases linearly with the number of relocators. The running time of the \textsc{OpR} model increases when reducing the maximum train size as the same relocation flow is split into more relocation tasks. However, the running times of the two considered benchmarks is orders of magnitude higher than our proposed operator-based scheme. In particular, the \textsc{AggR} case is particularly inefficient for a small relocation workforce. 

We recall that the \textsc{AggR} model is based on a complex combinatorial problem, and when there are less than 300~relocators the optimisation model is not able to find an idealised relocation plan that can serve all the potential demand. In this case, the search of the optimal relocation plan is a very complex task, and the time to solve the problem may grow exponentially with the size of the unserved demand. On the contrary, if an idealised relocation plan exists that ensures no dropped requests, the problem size reduces considerably. It is also useful to observe that the running time of \textsc{AR} in Fig.~\ref{s-fig:cputime1} corresponds to the computational time for solving Problem~\ref{pr:fluxes}. Consequently, the differences between the running time of \textsc{AR} and the running times of \textsc{OpR} is due to the computational time for solving Problem~\ref{pr:assignment}.

Finally, the results show that the running times of the \textsc{UB} model do not change significantly with the number of relocators, and they are comparable to the ones of the \textsc{AggR} model when there are more than 300 relocators. This is also due to the fact that, to cope with the computational complexity of the \textsc{UB} model, which  (differently from the \textsc{AggR} model) relocates both vehicles and drivers, we adopt the same aggregation approach for relocation variables as in~\cite{boyaci_optimization_2015}, and we relax the integer constraints.
%

%
%
%---------------------------------------------------

\subsection{Fleet utilisation \label{s-sec:utilisation}}
\begin{figure}[ht]
    \centering
    \includegraphics[width=0.8\columnwidth]{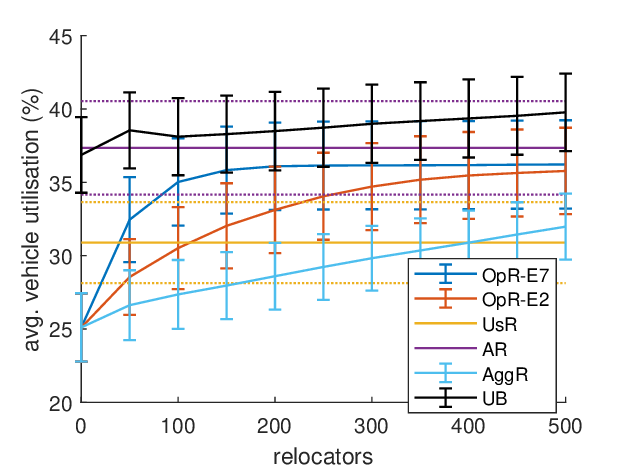}
    \caption{Average vehicle utilisation during the day for a fleet of 5,000 vehicles}
    \label{fig:util5k}
\end{figure}

\begin{figure}[ht]
    \centering
    \includegraphics[width=0.8\columnwidth]{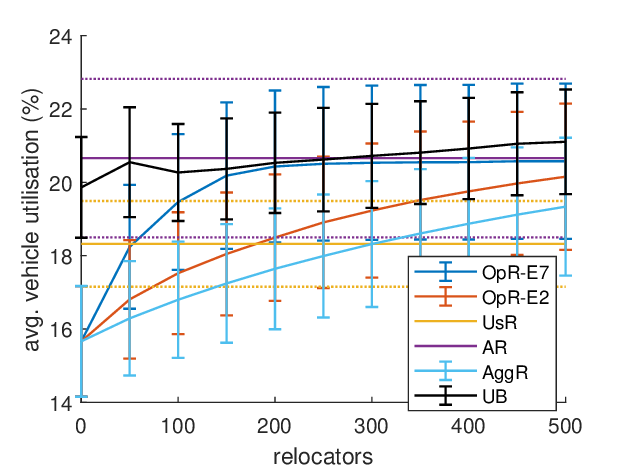} 
    \caption{Average vehicle utilisation during the day for a fleet of 10,000 vehicles}
    \label{fig:util10k}
\end{figure}

\noindent
\added{A critical performance metric for a car-sharing system is the utilisation of vehicles. In Figures~\ref{fig:util5k}--\ref{fig:util10k} we report the daily average vehicle utilisation for our model and for the benchmarks. We have calculated the fleet utilisation as the total vehicle-time spent with passengers during each day over the total vehicle-time available in a day. Since we assume the same demand for our model and the benchmarks, the utilisation is mainly a function of the dropped request rate and the fleet size. Therefore, for a given fleet size, the models that have lower dropped requests rates show higher fleet utilisation. It is also useful to point out that the maxi utilisation rate over a single time slot is higher than the average utilisation rate, since our demand is not constant throughout the day. For example, in the \textsc{OpR-E7} case with 200 relocators the daily peak utilisation reaches over 70\% and 40\% for a fleet of 5,000 and 10,000 vehicles, respectively. }

%
%---------------------------------------------------
\subsection{Effect of accuracy prediction for inventory imbalance \label{s-sec:imbalance}}
\begin{figure}[ht]
    \centering
    \includegraphics[width=0.8\columnwidth]{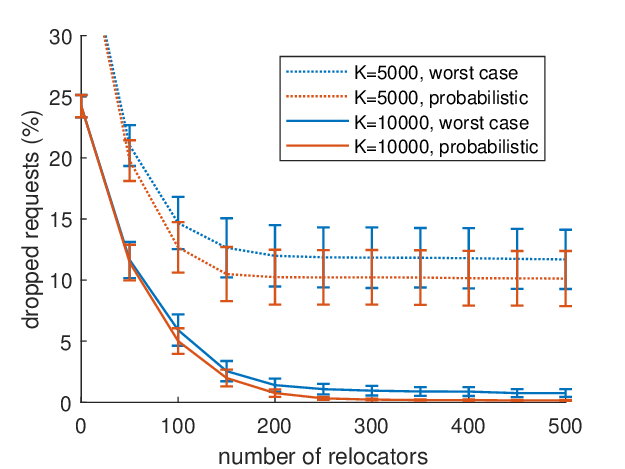}
    \caption{Percentage of rejected trip requests with the \textsc{OpR-E7} scheme for the worst-case and probabilistic estimate of inventory imbalance.}
    \label{s-fig:estimate_dropped}
\vspace{-0.2cm}
\end{figure}

\noindent
To conclude our evaluation, we analyse the effect on system performance of the two predictors we have designed in Section~\ref{sec:prediction}. The demand probability distributions $f^{k,t}_{A}(n;i)$ and $f^{k,t}_{C}(n;i)$ that are used in probability estimate of the inventory imbalance are extracted from the pickup and drop-off time series of the NYC taxi data using a simple histogram-method, which is based on counting the relative frequencies of the time series values
within each time slot. To this end, Fig.~\ref{s-fig:estimate_dropped} show the percentage of rejected trip requests with the \textsc{OpR-E7} when the inventory imbalance is estimated using the worst-case predictor or the probabilistic predictor. The results show that the probabilistic predictor ensures slightly better system performance for a given number of relocators. This is due to the fact that a more precise estimate of the surplus of vehicles at a feeder station is essential to select feasible relocation tasks. 
\bibliographystyle{ieeetr}
\bibliography{carsharing}